\let\dprod=\prod
\let\tprod=\prod
\let\tsum=\sum
\let\dsum=\sum
\let\text=\mbox

\documentclass[12pt]{article}
\usepackage{amssymb}
\usepackage[dvips]{epsfig}
\newtheorem{lemma}{Lemma}
\newtheorem{theorem}{Theorem}
\begin{document}
\def\thefootnote{\fnsymbol{footnote}}
\footnotetext{$^*$Supported in part by a grant 
from the Fulbright Foundation
and NSF.}         

\title{On Certain Two Dimensional Integrals that\\ Appear In 
Conformal Field Theory}
\author{Jeffrey S. Geronimo\footnotemark\\
School of Mathematics\\
Georgia Institute of Technology\\
Atlanta, GA  30332-0160
\and
Henri Navelet\\
CEA- Saclay\\
Service de Physique Theorique\\
F-91191 Gif-sur-Yvette Cedex, France}
\maketitle
\section{Introduction}

In the framework of perturbative QCD at large $Q^{2}$ and low $x$, the
conformal (global) group of transformations in the transverse coordinate
space plays a crucial role.  Indeed Lipatov and his collaborators [1] have
derived the master equation for the derivative of the gluon structure
function with respect to $x$ by resuming the leading $\left( \alpha
_{s}\left( \bar{Q}^{2}\right) \log (1/x)\right) ^{n}$ terms at fixed $\alpha
_{s}\left( \bar{Q}^{2}\right) .$ This derivative is a convolution in the
transverse space of the gluon structure function times a conformal invariant
kernel (BKKL kernel). All the relevant observables are then expressed as an
expansion over the basis $E^{n,\nu }$ of the conformal eigenfunctions of
this kernel. The integer $n$ is the conformal spin and $i\nu $ corresponds
to a continuous imaginary scaling dimension. Among others, the elastic off
mass shell gluon-gluon amplitude corresponding to the exchange of the bare
QCD hard Pomeron [1, 2] and the conformal invariant triple Pomeron vertex
[3]. For this purpose one is led following [1] to consider the 
eigenfunctions in a mixed representation which is obtained by a 
Fourier transform in 2 dimensions of $E^{n,\nu }.$ To be more specific, 
these eigenfunctions
in the coordinate space are the product of a holomorphic times an
antiholomorphic function. Recently, it has been shown [2] that the answer is
a sum of two products of a holomorphic times an antiholomorphic form. This was accomplished by
noticing that the Fourier transform is solution of a set of two linear
differential equations one with respect to the complex variable and the
other with respect to the complex conjugate variable. This structure of
conformal blocks is already present in the computation of correlation
functions in 2 dimensional conformal invariant quantum field theories [4].
Various theorems have been established for an integrand with integrable
singularities at $0,\;1$ and $\infty $. In reference [2] a similar 
result seems to apply, as shown by the existence in this case of 2
differential equations. In the first part of this paper we show that
this is not a sheer coincidence but that similar theorems can apply for the
bidimensional Fourier transform. As far as the triple conformal invariant
Pomeron is concerned, specific analytic calculations have been done recently
[3]. It turns out that this quantity is a peculiar case of a more general
calculation involving a p-tuple conformally invariant integral in the complex
space at a particular complex value of $z$ (resp $\bar{z}).$ The triple Pomeron
corresponds to the case $z=1$ and $p=3.$ Our second part is the derivation
of this very general result. The key point is to find the set of two linear
differential equations with respect to $z$ and $\bar{z}$ of order $p+1$
which are obeyed by the p-tuple integral. The final answer will thus be a sum
of conformal blocks, as expected, each block being the product of one of the 
$p+1$ solutions in $z$ of the differential equation times the corresponding
one in $\bar{z},$ in order to preserve the single valuedness of the
solution. In our study, the system of linear equations is the one obeyed by
the hypergeometric function $_{p+1}\!F_{p}(z).$ The case of
degeneracy which is relevant for the calculation of the triple
Pomeron vertex will be treated explicitly in a forthcoming
paper.

\section{Bidimensional Fourier Transforms}

We begin this section with the following,

\begin{lemma} For $0\le x\le y\le{\pi\over 2}$, let
$$
I_1 = \int_0^{\pi\over 2}\left( 
\int_0^y e^{-R( \sin y - \sin x)}dx\right)dy.
$$
Then $|I_1|\le C_{\epsilon}R^{\epsilon -1}$,
for any $0<\epsilon\le1$. 
\end{lemma}

\noindent Proof. It  follows from the mean value Theorem and
Jordan's lemma that for $0\le x\le y\le{\pi\over 2},\  \sin y - \sin x \ge
{\pi\over 2}({\pi\over 2} -y)(y-x)$, consequently,
$$
\int_0^y e^{-R( \sin y - \sin x)}dx \le 
\int_0^y e^{-R{\pi\over 2}({\pi\over 2} -y)(y-x)}dx = 
{\pi\over 2R}{1- e^{-{2R\over\pi}({\pi\over 2} -y)y}\over({\pi\over 2} -y)}.
$$
Substituting the above result into $I_1$, then integrating by parts yields
$$
I_1 \leq  \int_0^{\pi\over 2}\ln({\pi\over 2} -y)
e^{-{2R\over\pi}({\pi\over 2} -y)y}({\pi\over 2} -2y)dy.
$$
Since $|({\pi\over 2} -y)^{\epsilon}\ln({\pi\over 2} -y)|
\le K_{\epsilon} < \infty$
for $0\le y\le{\pi\over 2},\ \epsilon > 0$ with $K_{\epsilon}$
monotonically decreasing  we see that
\begin{eqnarray*}
I_1 &\le &{\pi\over 2} K_{\epsilon}\int_0^{\pi\over 2}
({\pi\over 2} -y)^{-\epsilon}e^{-{2R\over\pi}({\pi\over 2} -y)y}dy\\
    &\le&{\pi\over 2}K_{\epsilon}({\pi\over 4})^{-\epsilon}
\int_0^{\pi\over 4}e^{-({R\over2}y)}dy +  
{\pi\over 2}K_{\epsilon}\int_{\pi\over 2}^{\pi\over 4}
({\pi\over 2} -y)^{-\epsilon}e^{-{2R\over\pi}({\pi\over 2} -y)y}dy.
\end{eqnarray*}
Thus extending the region of integration in the first integral to 
the full semi-axis then doing  same thing to the second integral after 
the change of variables $y_1= {\pi\over 2} -y$ yields,
$$
I_1\le {\pi\over 2}({\pi\over 4})^{-\epsilon}
{2K_{\epsilon}\over R}+{\pi\over 2}K_{\epsilon}
({2\over R})^{1-\epsilon}\Gamma(1-\epsilon).
$$
which gives the result. 

With the above lemma we can now factorize certain double integrals.  We will 
consider integrals of the form
$$I=\mathop{\int\!\int}_{{\mathbb C}\backslash [0,\infty)} f(t)
\overline{g(t)}\, d^2t,$$
where ${\mathbb C}\backslash [0,\infty)$ is the complex plane cut along the real
axis from zero to infinity.  We will make the following assumptions
about $f$ and $g$.

\begin{itemize}
\item[i)] $I$ converges.
\item[ii)] $f(t)=\tilde f(t)e^{iq\cdot t}$, $g(t)=\tilde g(t)e^{iq\cdot t}$,
$q\in{\mathbb R}$.
\item[iii)] $\tilde f(t)$ and $\tilde g(t)$ are analytic for $t\in
{\mathbb C} \backslash [0,\infty)$.
\item[iv)] $k_f=\left|\int^\infty_0 f(x)dx\right|<\infty$,
$k_g=\left|\int^\infty_0 g(x)dx\right|<\infty$.
\item[v)] For $\epsilon$ small enough but positive 
$\lim_{R\to\infty} k_{\tilde g} (R)R^\epsilon=0$,
$\lim_{R\to\infty} k_{\tilde f} (R)R^\epsilon=0$ and
$\lim_{R\to\infty} k_{\tilde g} (R)k_{\tilde f}(R)
R^{1+\epsilon}=0$, where 
$k_{\tilde g}(R)=\max_{\theta\in [0,2\pi)}
|\tilde g(Re^{i\theta})|$ and\newline
$k_{\tilde f}(R)=\max_{\theta\in [0,2\pi]}
|\tilde f(Re^{i\theta})|$. Finally
\item[vi)] $\lim_{r\to 0} k_{\tilde f}(r)r=0=\lim_{r\to 0}k_{\tilde g}
(r)r$.
\end{itemize}

With the above assumptions $I=\lim_{R\to\infty} I_R$ where
$$I_R=\mathop{\int\!\!\int}_{D_R\backslash [0,R)}
f(t)\overline{g(t)}\,d^2t$$
with $D_R$ being the disk of radius $R$.
We proceed to cut $D_R\backslash [0,\infty)$ into the components indicated
in Figure~1 below.
If we choose the point $z_1$ as the base point then Stoke's Theorem 
implies that
$$I_R=\lim_{\Delta\to 0} {i\over 2}\int_{\partial D_1+\partial D_2}
f_1(t)\overline{g(t)}\, \overline{dt}$$
where $\partial D_i$, $i=1,2$ indicate the boundary of $D_i$, $i=1,2$
respectively, $f_1(t)=\int_{\Gamma_{1,t}}f(y)dy$,
$\Gamma_{1,t}$ being the path followed from the point $z_1$ to $t$
along the contour indicated on $\partial D_1$ or $\partial D_2$.
Note that the contributions to $I_R$ along $\Gamma_{8541}$ on 
$\partial D_2$ cancels with the contribution of $\Gamma_{1458}$ on
$\partial D_1$. The integral
\begin{eqnarray*}
\left|\int_{\Gamma_{21}}f_1(t)\overline{g(t)}\,\overline{dt}\right|&=&
\left|\int^{\pi/2}_0d\phi \int^{\pi/2}_\phi d\theta
f(Re^{i\theta})\overline{g(Re^{i\phi})}\, e^{i(\theta-\phi)}R^2\right|\\
&\le& k_{\tilde g}(R)k_{\tilde f}(R)R^2\int^{\pi/2}_0
d\phi \int^{\pi/2}_\phi d\theta e^{-qR(\sin\theta-\sin\phi)}
\end{eqnarray*}
which by Lemma~1 is bounded by
\begin{equation}\label{eq1}
\le k_{\tilde g}(R)k_{\tilde f}(R)R^{1+\epsilon}.
\end{equation}
$\epsilon$ small but positive.  Since
$$\int_{\Gamma_{18}} f_1(t)\overline{g(t)}\,\overline{dt}=
\int^{3\pi/2}_{\pi/2}d\phi \int^\phi_{\pi/2}
d\theta f(Re^{i\theta})\overline{g(Re^{i\phi})} e^{i(\theta-\phi)}
R^2$$
An argument similar to the one above results in a bound similar to
(\ref{eq1}). The integral
\begin{eqnarray*}
&&\left|\int_{\Gamma_{87}}f_1(t)\overline{g(t)}\,\overline{dt}\right|=\\
&&\quad \left|\int_{\Gamma_{87}}(f_7+f_6-f_7+f_3-f_6+f_2-f_3+f_1-f_2)
\overline{g(t)}\overline{dt}\right|
\end{eqnarray*}
and is bounded by
$$\le 2k_{\tilde g}(R)k_{\tilde f}(R)R^{1+\epsilon}+
2k_fk_{\tilde g}
R^\epsilon+rk_{\tilde f}(r)k_{\tilde g}(R)R^\epsilon$$
The above inequalities also show that
$$\lim_{\Delta\to 0}\left| \int_{\Gamma_{63}} f_1(t)\overline{g(t)}dt\right|=0;
$$
$$\left|\int_{\gamma_{32}} f_1(t)\overline{g(t)}\,\overline{dt}-
\int_{\Gamma_{32}}f_2\overline{g(t)}\,\overline{dt}\right|\le k_gk_{\tilde f}
R^\epsilon$$
and
$$\lim_{\Delta\to 0}\left|\int_{\Gamma_{76}}
f_1(t)\overline{g(t)}\,\overline{dt}- \left((f_2-f_3)+(f_6-f_7)
+f_7(t)\right)\overline{g(t)}\,\overline{dt}\right|\le
k_g k_{\tilde f}R^\epsilon_+$$
This leads to,

\begin{lemma} Suppose $f$ and $g$ satisfy {\rm i)--vi)} then
\begin{eqnarray*}
&&\lim_{R\to\infty}\Bigl| I_R-\lim_{\Delta \to 0}
\biggl({i\over 2}\int_{\Gamma_{76}}\left((f_2-f_3)+(f_6-f_7)\right)
\overline{g(t)}\,\overline{dt}\\
&&\quad + {i\over 2}\int_{\Gamma_{32}}f_2(t)\overline{ g(t)}\,\overline{dt}
+{i\over 2}\int_{\Gamma_{76}} f_7(t)\overline{g(t)}\,\overline{dt})\biggr)\Bigr|
=0
\end{eqnarray*}
\end{lemma}

We now consider the integral
$$I=\mathop{\int\int}_{\hat C}
t^{\gamma-1}\bar t^{\tilde\gamma-1}e^{i(\bar q\cdot t+ q\cdot \bar t)}
d^2t.$$
where $\hat C$ is the complex plane cut so as to make $\ln t$ well defined.
By rotating and scaling the coordinate system we may write
$$
I = q^{-\tilde\gamma}{\bar q}^{-\gamma} \hat I,
$$
where 
$$I=\mathop{\int\int}_{{\mathbb C}\backslash [0,\infty)}
t^{\gamma-1}\bar t^{\tilde\gamma-1}e^{i(t+\bar t)}
d^2t.$$ 
In the above integral the branch of the logarithm selected is such that 
$0\le \arg t
< 2\pi$ and $\ln t$ is positive for $t>1$.  With $t=re^{i\theta}$ set
$$f(t)=t^{\gamma-1} e^{ir\cos\theta-r\sin\theta}$$
and
$$g(t)=t^{\bar{\tilde \gamma}-1}e^{-ir\cos\theta+r\sin\theta}.$$
Thus we find
$$\hat I=\int^\infty_0 \int^{2\pi}_0 r^{\gamma+\tilde \gamma-1}
e^{i(\gamma-\tilde \gamma)\theta} e^{2ri\cos\theta} dr\,d\theta\qquad
q>0.$$
Since $\int^{2\pi}_0 e^{i(\gamma-\tilde\gamma)\theta} e^{2ir\cos\theta}
d\theta=O\left(1\over\sqrt r\right)$ we see that $\hat I$ converges uniformly
for $0<{\rm Re}(\gamma+\tilde\gamma)<1/2$ and thus defines an analytic function
of $\gamma$ and $\tilde\gamma$ for the values of these variables restricted
to this domain.

We now consider the integral
$$I_R=\mathop{\int\int}_{D_R\backslash [0,R]} f(t)\overline{g(t)}
d^2t$$
with $0<{\rm Re}\; \gamma <1/2$, $0<{\rm Re}\, \tilde\gamma <1/2$ and
$0<{\rm Re}(\gamma+\tilde \gamma)<1/2$.  If $f_1(t)=t^{\gamma-1}$
and $g_1(t)=t^{\bar{\tilde \gamma}-1}$  the conditions
of Lemma~2 are satisfied and  taking into account the increment in the phase
due to the cut we find
\begin{eqnarray*}
\hat I&=& \lim_{R\to\infty} I_R={1\over 2}\left(1-e^{2\pi i\gamma}\right)
e^{-2\pi i\tilde\gamma}\int^\infty_0 x^{\gamma-1}
e^{ix}dx\int^\infty_0 y^{\tilde \gamma-1}e^{iy}dy\\
&&+{1\over 2}\left( e^{2\pi i(\gamma-\tilde
\gamma)}-1\right)I(\gamma,\tilde\gamma),
\end{eqnarray*}
where 
$$I(\gamma,\tilde\gamma) = \int^\infty_0
dv\int^\infty_v du u^{\gamma-1}v^{\tilde \gamma-1}e^{i(u+v)}.
$$
With the first two integrals replaced by their representation 
[6, Formula (33), p.~12]
in terms of gamma functions we find
$$\hat I={1\over 2}(1-e^{2\pi i\gamma})e^{-2\pi i\tilde\gamma}
e^{i(\gamma+\tilde\gamma)\pi/2}
\Gamma (\gamma)\Gamma(\tilde\gamma)+{1\over 2} 
(e^{2\pi i(\gamma-\tilde\gamma)}-1)I_1.$$
The first term can be analytically extended to $\gamma,\tilde\gamma
\ne 0,-1,-2,\dots$. To solve the 2nd term we break it up into three
integrals over the following regions $R_1=\{(v,u), 0\le v<1, 
v\le u<1\}$, $R_2=\{(u,v), 0\le v<1, 1\le u\le \infty\}$ and
$R_3=\{(u,v), 1\le v<\infty, v\le u<\infty\}$.
It is easy to see that $I_1$ can be analytically extended to 
$\gamma\ne 0,-1,-2,\dots$, $\gamma+\tilde\gamma\ne 0,-1,-2,\dots$ and
$I_2=\int^1_0 v^{\tilde\gamma} e^{iv}du\int^\infty_1
u^{\gamma-1}e^{iu}du$ can be extended to the region Re~$\gamma<1$,
$\tilde\gamma\ne 0,-1,-2,\dots$.
For Re~$\gamma<1$, 
\begin{eqnarray*}
|I_3|&=&\left|\int^\infty_1 v^{\tilde\gamma-1} e^{iv}
\int^\infty_v u^{\gamma-1}e^{iu}dudv\right|\\
&\le&C\int^\infty_1 v^{{\rm Re}(\tilde\gamma+\gamma)-2}dv
\end{eqnarray*}
which converges uniformly for Re~$(\gamma+\tilde\gamma)<1$.
Consequently $I_3$ can be extended to Re~$\gamma <1$, 
Re$(\gamma+\tilde\gamma)<1$. This implies that $I$ can be extended toe the region Re$(\gamma)<1$, Re$(\gamma+\tilde\gamma)<1$, $\gamma, \tilde\gamma\ne
0,-1,-2,\dots$.  For Re~$\gamma >0$ $I(\gamma,\tilde\gamma)$
 can
be written as
\begin{eqnarray*}
I(\gamma,\tilde\gamma) &=&\int^\infty_0 \int ^\infty_0 u^{\gamma-1} v^{\tilde\gamma-1}
e^{i(u+v)}du\,dv\\
&&-\int^\infty_0 \int^v_0 u^{\gamma-1} v^{\tilde\gamma-1}
e^{iq(u+v)}du\,dv .
\end{eqnarray*}
which allows an analytic extension for Re~$\gamma >0$, $0<{\rm Re}(\gamma
+\tilde \gamma)<1$ since this implies that ${\rm Re}\tilde\gamma<1$.  Consequently we have shown

\begin{theorem}
For $0< {\rm Re}(\gamma+\tilde\gamma)<1$, $\gamma, \tilde\gamma\ne
0,-1,-2,\dots$, $q\ne 0$ 
\begin{eqnarray*}
I&=&{i\over 2}(1-e^{2\pi i\gamma})e^{-2\pi i\tilde\gamma}
q^{-\tilde\gamma}{\bar q}^{-\gamma} e^{i(\gamma+\tilde\gamma)\pi/2}
\Gamma(\gamma)\Gamma(\tilde\gamma)\\
&&+{i\over 2}(e^{2\pi i(\gamma-\tilde\gamma)}-1)I(\gamma,\tilde\gamma) .
\end{eqnarray*}
\end{theorem}

In particular if $\gamma-\tilde\gamma=n$, $n$ an integer the integral becomes
$$I=\sin(\pi\gamma)e^{i(\pi/2)n}q^{-\tilde\gamma}{\bar q}^{-\gamma} \Gamma(\gamma)
\Gamma (\gamma+n)$$

\begin{figure}
\centerline{\epsfig{figure=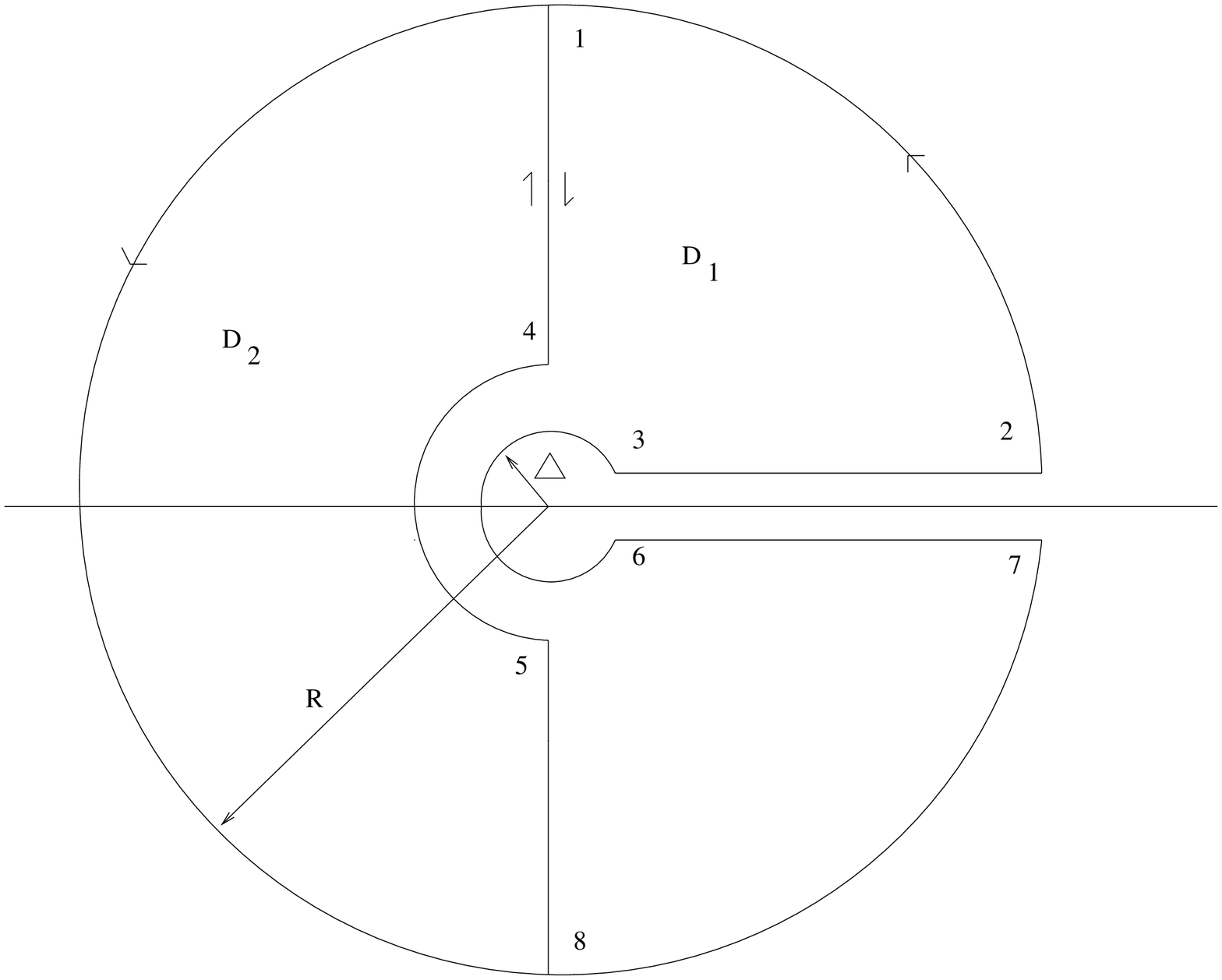,height=5in,width=6in}}
\caption{}
\end{figure}

The next integral we will consider arises in the context of the 
QCD Domain. The integral to be considered is
$$I=\int\int d^2z\ {e^{i/2(q\bar z+\bar q z)}
\over\left(z^2-{\rho^2\over 4}\right)^{u+1/2}
\left(\bar z^2-{\bar\rho^2\over 4}\right)^{\hat u+1/2}},
$$
 where $u=-v_1+iv_2,\ \hat u=v_1+iv_2$, and  Re$(q\bar\rho)\ne0$. 
With an appropriate rotation and scaling the above integral becomes
$$I={1\over \left(\rho^2\over 4\right)^u \left(\bar\rho^2
\over 4\right)^{\hat u}}I_1$$
with 
$$I_1=\mathop{\int\!\!\int}_{\kern-10pt\hat C}{d^2z\ e^{(q_1\cdot\bar z
+\bar q_1\cdot z)}\over (z^2-1)^{u+1/2}(\bar z^2-1)^{\hat u+1/2}},$$
and $q_1=q{\bar\rho\over 4}$. Since $q_1\bar z$ comes in symmetrically with
its complex conjugate we may assume that  ${\rm Re} q_1 >0$ and write
$q_1=\hat q_1 e^{i\phi}$. With this notation  $\hat
{\mathbb C}$ in the above integral is the complex
plane cut from $[1,\infty e^{i\phi} )$ and from $(-\infty e^{-i\phi},-1]$. We use the branch
of the logarithm so that the phases of $z-1$ and $z+1$ are equal to zero
for $z > 1$.  Elementary methods including
stationary phase shows that
the above integral is convergent and defines an analytic
function in the variables $u$ and $\hat u$  for $-1/4<
{\rm Re}(u+\hat u)<1$ and ${\rm Re} q_1 \ne 0$ which is also continuous in 
$q_1$ in this region. 
We restrict $u$ and $\hat u$ to the region $-1/4<{\rm Re}\; u, {\rm Re}\;
\hat u <1/4$ and consider the integral
$$I_R=\mathop{\int\!\!\int}_{\kern-10pt\hat D_R}
 {d^2z e^{1/2(\bar q_1\cdot z
+q_1\cdot \bar z)}\over (z^2-1)^{u+1/2}(\bar z^2-1)^{\hat u+1/2}}$$
Using Stoke's Theorem we integrate over $\partial D_1$ and
$\partial D_2$ given in Figure~2 with $\tilde f=(z^2-1)^{-(u+1/2)}$
and $\tilde g=(z^2-1)^{\bar{\hat u}+1/2}$ to find,
$$I_R= {i\over 2}\oint_{\partial D_1+\partial D_2} f_1(z)
\overline{g(z)}dz.$$

The conditions on $u$ and $\hat u$ are such that hypotheses i - iii,v, and
vi of Lemma~2 follow almost immediately. To verify iv) for the above functions
we begin by showing that
$|\int_{\Gamma_{23}}\overline{g(z)}d\bar z|< \infty$. Thus we need  to show
that
$\lim_{R\to\infty}|\int_0^{r_0} \overline{g(z)}d\bar z|<\infty$ with
$z=1+re^{i\hat \phi}$, and with $r_0$ and $\hat\phi$ determined by the
equation $1+r_0e^{i\hat\phi} = Re^{i\phi}$. The conditions on $r_0$ and
$\hat\phi$ imply that 
$r_0=R(1+ o(1))$ and $\hat\phi=\phi + O({1\over R})$. Now 
$$
|\int_0^b \bar g d\bar z|< C_1 e^{\hat q_1 b \sin(\hat \phi - \phi)}\int_0^b
r^{-{\rm Re}\hat u - {1\over2}}dr < \infty,\ {\rm Re} \hat u < 1/2,
$$
and using integration by parts,
$$
|\int_b^{r_0}\overline{g(z)} d\bar z|< C_2 {e^{\hat q_1 \sin(\hat \phi -
\phi)}\over |i\cos(\hat \phi -\phi)+\sin(\hat \phi -\phi)|}{b^{-(2{\rm Re}\hat
u +1)}\over |q|} < \infty,\ -{1\over 2}<{\rm Re} \hat u,
$$
for constants $C_1$ and $C_2$ independent of $R$. Applying the above
reasoning to $f$ on $\Gamma_{45}$ shows that for $-1/4<{\rm Re}\; u, {\rm Re}\;
\hat u <1/4$, iv is satisfied. Thus Lemma~2 (appropriately modified) shows that 
\begin{eqnarray*}
I&=&\lim_{R\to\infty}{i\over 2}\left[\int_{\Gamma_{54}}
(f_2-f_5)\overline{g(z)} d\bar z+\int_{\Gamma_{54}}
f_5\overline{g(z)} d\bar z+\int_{\Gamma_{32}}f_2\overline{
g(z)}d\bar z\right]\nonumber\\
&&+\lim_{R\to\infty}{i\over 2}\left[\int_{\Gamma_{87}}
(f_{10}-f_7)\overline{g(z)} d\bar z+
\int_{\Gamma_{87}} f_7 \overline{g(z)}d\bar z+\int_{\Gamma_{109}}
f_{10} \overline{g(z)}d\bar z\right]
\end{eqnarray*}
  From [7, p.~167 eq (6)]   we see that 
$$\lim_{R\to\infty}(f_5-f_2) =
{i\pi\Gamma({1\over2})({\bar q_1\over2})^u\over\Gamma(u+{1\over2})}H^{(1)}_{-u}(\bar q_1),
$$
and
$$
\lim_{R\to\infty}(f_{10}-f_7)={i\pi\Gamma({1\over2})({\bar
q_1\over2})^u\over\Gamma(u+{1\over2})}H^{(2)}_{-u}(\bar q_1).
$$ 
Furthermore for $\hat u$ restricted to the region of interest the radius of
small circular regions around $\pm1$ may be taken to zero so that

\begin{eqnarray*}
\lim_{R\to \infty}\int_{\Gamma_{54}}\overline{g(z)} d\bar z&
=&-\int_0^{\infty}{e^{i\hat q_1e^{i\phi}(1+re^{-i\phi})}e^{-i\phi}\over
(re^{-i\phi})^{\hat u+{1\over2}}(2+re^{-i\phi})^{\hat u+{1\over2}}}d r\cr
&=&{i\pi\Gamma({1\over2})({q_1\over2})^{\hat u}\over2\Gamma({1\over2}-\hat
u)}H^{(1)}_{\hat u}(q_1),
\end{eqnarray*}  
and
$$
\lim_{R\to \infty}\int_{\Gamma_{87}}\overline{g(z)} d\bar z
={i\pi\Gamma({1\over2})({q_1\over2})^{\hat u}\over2\Gamma({1\over2}-\hat
u)}H^{(2)}_{\hat u}(q_1).
$$
Thus
\begin{eqnarray}
I_1&=&{\pi\Gamma(1/2-\hat u)\over4\Gamma(1/2+u)}
\left(q_1\over 2\right)^{\hat u}\left(\bar q_1\over 2\right)^u\nonumber\\
&&\times [H^1_{-u}(\bar q_1)H^1_{\hat u}
(q_1)-H^2_{-u}(\bar q_1)H^2_{\hat u}(q_1)]\nonumber \\
&& +2i\sin\pi (\hat u-u)\sin[(\pi-\bar q_1-q_1)(\hat u-u)]e^{i\phi(\hat u-u)} I(\hat q_1),\ q_1\ne0\end{eqnarray}
where
$$I(\hat q_1)=\int^\infty_0 \int^\infty_y {e^{i\hat q_1(r + y)}\over
(r(2+re^{i\phi}))^{u+1/2}(y(2+ye^{-i\phi}))^{\hat u+1/2}}dr\,dy.$$
Using an argument similar to the one given above but more tedious
it can be shown that $I(\hat q_1)$ has an analytic
extension $-1/4 < {\rm Re}\, u+{\rm Re}\,\hat u < 1$,
$u,\hat u\ne 1/2+ n, n= 0,1,...$
Thus we have shown 

\begin{theorem} For $u=-v_1+iv_2$, $\hat u=v_1+iv_2$,
 $u,\hat u\ne 1/2 + n, n = 0,1,...$, $I_1$ is given by equation $(2)$.
In particular if $v_1=n/2$, $v_2\ne 0$ then
\begin{eqnarray}
I_1&=& {\pi\Gamma(1/2-\hat u)\over 4\Gamma(1/2+u)}
\left(q_1\over 2\right)^{\hat u}\left(\bar q_1\over 2\right)^u\nonumber \\
&& [H^1_{-u} (\bar q_1)H^1_{\hat u} (q_1)- H^2_{-u}
(\bar q_1) H^2_{\hat u} (q_1)]\qquad q_1\ne0.
\end{eqnarray}
\end{theorem}

\begin{figure}
\centerline{\epsfig{figure=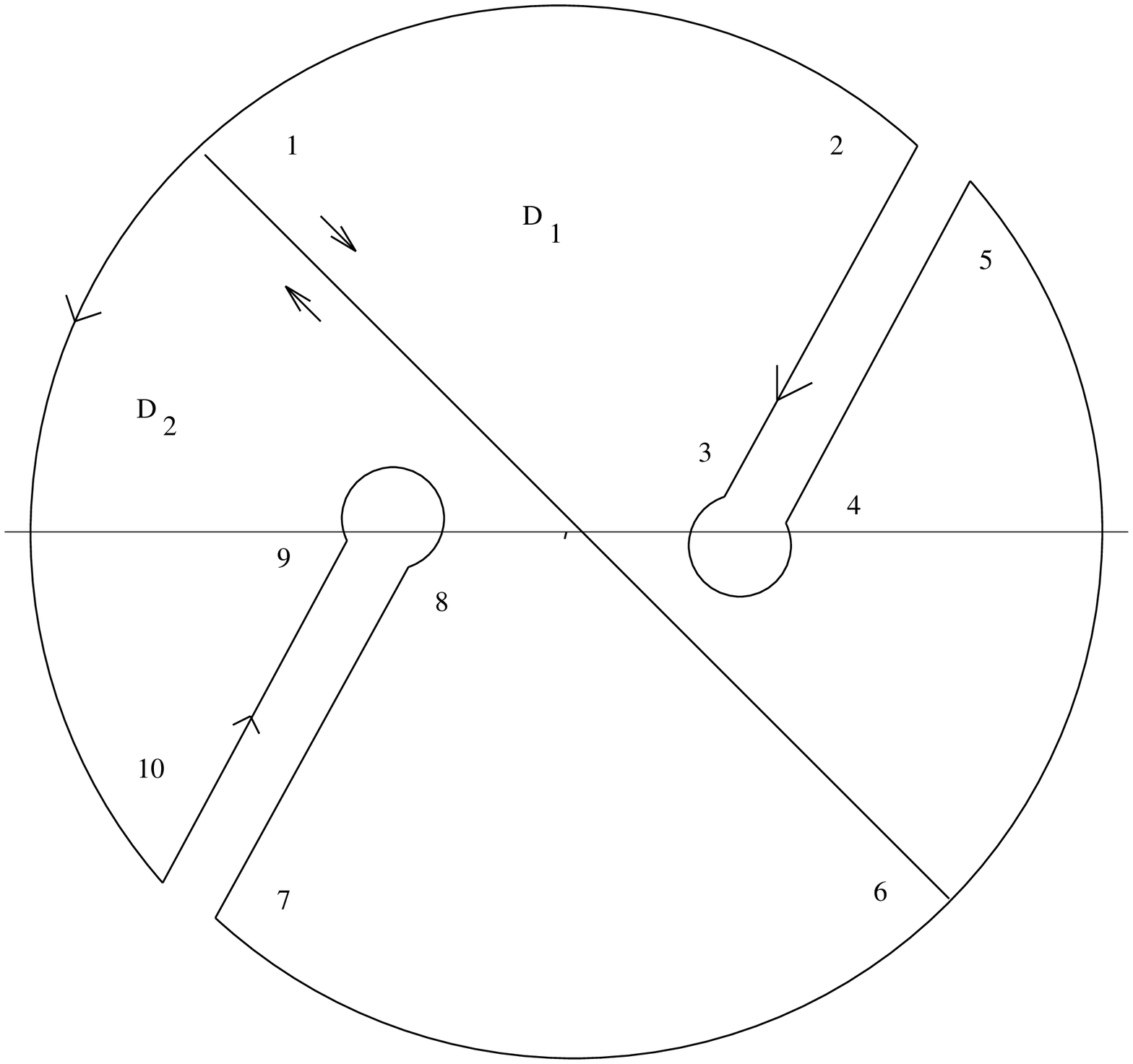,height=5in,width=5in}}
\caption{}
\end{figure}

\section{A p-uple conformal integral and applications}
We now consider the integral

\begin{eqnarray}
&&I_{p+1}\left( a_{0},\;a_{i},\;b_{i};\;\tilde{a}_{0},\;\tilde{a}_{i},\;%
\tilde{b}_{i};\;z,\bar{z}\right)  \nonumber \\
&=&\left( \frac{1}{2i}\right) ^{p}\int \dprod\limits_{i=1}^{p}d^2z_{i}\;\left(
z_{i}\right) ^{a_{i}-1}\left( \bar{z}_{i}\right) ^{%
\tilde{a}_{i}-1}\left( 1-z_{i}\right) ^{b_{i}-a_{i}-1}\left( 1-\bar{z}_{i}\right) ^{\tilde{b}_{i}-\tilde{a}_{i}-1}\nonumber \\ & &\quad \left(1-\left( \prod\limits_{i=1}^{p}z_{i}\right) z\right) ^{-a_{0}}\left( 1-\left( \prod\limits_{i=1}^{p}\bar{z}_{i}\right) \bar{z}%
\right) ^{-\tilde{a}_{0}}  
\end{eqnarray}
We will analyze the case when $a_{i},\tilde a_{i}, b_{i}$ and $\tilde b_{i}$ obey the following conditions:
\bigskip

\noindent{\bf Condition C}. For $ i = 0\ldots, p$ and $ j = 1\ldots, p$,

\begin{itemize}
\item[a)] $b_{j}, \tilde b_{j}, a_{i}, \tilde a_{i} \not\in {\bf Z}.$
\item[b)] $a_{i}-a_{j}, \tilde a_{i}- \tilde a_{j}, b_{i\ne 0}-b_{j}, \tilde b_{i\ne 0}- \tilde b_{j}, \ i\ne j\ \not\in {\bf Z}.$
\item[c)] $a_{i}-b_{j},\tilde a_{i}- \tilde b_{j} \not\in {\bf Z}$
\item[d)] $a_{i}- \tilde a_{i}, b_{j}-\tilde b_{j} 
\in {\bf Z}.$ 
\end{itemize}

With $a^j =(a_1,\ldots,a_j)$, $b^j =(b_1,\ldots,b_j)$, $\tilde a^j =(\tilde a_1,\ldots,\tilde a_j)$, $\tilde b^j =(\tilde b_1,\ldots,\tilde b_j)$, and $ I_1(a_0, \tilde a_0,z,\bar z) = (1- z)^{-a_0}(1- \bar z)^{-\tilde a_0}$ the above integral can be recast as the iterated integral

\begin{eqnarray}
&&I_{p+1}\left( a_{0},\;a^p,\;b^p;\;\tilde{a}_{0},\;\tilde{a}^p,\;%
\tilde{b}^p;\;z,\bar{z}\right)  \nonumber \\
&=\left( \frac{1}{2i}\right)& \int d^2z_{p} \left(
z_{p}\right) ^{a_{p}-1}\left( 1-z_{p}\right)^{b_{p}-a_{p}-1}%
\left(\bar{z}_{p}\right)^{\tilde{a}_{p}-1}\left( 1-\bar{z}_{p}\right)%
^{\tilde{b}_{p}-\tilde{a}_{p}-1}\nonumber \\
&& I_{p}\left( a_{0},\;a^{p-1},\;b^{p-1};\;%
\tilde{a}_{0},\;\tilde{a}^{p-1},\tilde{b}^{p-1};\;z\;z_p,\bar{z}\;\bar{z_p}\right),
\end{eqnarray}

The computation of this integral has already been done for 
$p=0$ and $p=1$ by various
authors [4] in the case when $a_{i}=\tilde{a}_{i},\;b_{i}=\tilde{b}%
_{i}$ and $a_{0}=\tilde{a}_{0}.$ The result is a sum of a product of
hypergeometric functions of $z$ with its antiholomorphic part of argument $%
\bar{z}$ exhibiting the conformal block structure. Lipatov [1] has given the
general form for $p=1$ without the restrictions on the $\tilde{a}_{i},\;%
\tilde{a}_{0},\;\tilde{b}_{i}.$

We propose  to give the exact analytic structure of  $%
I_{p+1}$ for any positive integer value of $p.$

The key point of the calculation is to prove that the $I_{p+1}$ obey two linear
differential equations of order $p+1$ namely

\[
O_{z}^{p+1}\left( a_{0},\;a^{p},\;b^{p}\right) I_{p+1}\left(
a_{0},\;a^{p},\;b^{p};\;\tilde{a}_{0},\;\tilde{a}^{p},\;\tilde{b}^{p};\;z,\;%
\bar{z}\right) \equiv 0 
\]

\noindent and

\[
O_{\bar{z}}^{p+1}\left( \tilde{a}_{0},\;\tilde{a}^{p},\;\tilde{b}^{p}\right)
I_{p+1}\left( a_{0},\;a^{p},\;b^{p};\;\tilde{a}_{0},\;\tilde{a}^{p},\;\tilde{%
b}^{p};\;z,\;\bar{z}\right) \equiv 0 
\]

\noindent where $O_{z}^{p+1}\left( \text{resp. }O_{\bar{z}}^{p+1}\right) $
is the differential operator of order $p+1$ defining the hypergeometric
function $_{p+1}\!F_{p}\left( a_{0},\;a^{p},\;b^{p},\;z\right) $ $\left( 
\text{{\rm (resp.)}}{\rm \ }_{p+1}\!F_{p}\left( \tilde{a}_{0},\;\tilde{a}%
^{p},\;\tilde{b}^{p},\;\bar{z}\right) \right) $

\noindent namely 
\[
O_{z}^{p+1}\equiv \frac{\partial}{\partial z}\prod\limits_{i=1}^{p}
\left( z\frac{\partial}{\partial z}%
+b_{i}-1\right) -\prod\limits_{i=0}^{p}\left( z\frac{\partial}{\partial z}
+a_{i}\right) 
\]

resp.

\begin{equation}
O_{\bar{z}}^{p+1}\equiv \frac{\partial}{\partial\bar{z}}
\prod\limits_{i=1}^{p}\left( \bar{z%
}\frac{\partial}{\partial\bar{z}}+\tilde{b}_{i}-1\right) 
-\prod\limits_{i=0}^{p}\left( 
\bar{z}\frac{\partial}{\partial\bar{z}}+\tilde{a}_{i}\right)  
\end{equation}

The general solution of this system of differential equation reads
\[
I_{p+1}\left( z,\;\bar{z}\right) =\tsum\limits_{i,j=0}^{p}\;\lambda
_{ij}^{\left( p+1\right) }\;u_{i}^{p+1}\;\left( z\right) \;\tilde{u}%
_{j}^{p+1}\;\left( \bar{z}\right) 
\]

\noindent where $u_{i}^{p+1}\left( a_{0},\;a^{p},\;b^{p},\;z\right) $ $%
\left( {\rm resp.\ }\tilde{u}_{j}^{p+1}\left( \tilde{a}_{0},\;\tilde{a}%
^{p},\;\tilde{b}^{p},\;\bar{z}\right) \right) $ is any of the $p+1$
independent solutions of the differential equation

\[
O_{z}^{p+1}\varphi _{j}\left( z\right) =0\qquad j=0,..,\;p 
\]

\noindent namely

\[
u_{0}^{p+1}\left( a_{0},\;a^{p},\;b^{p},\;z\right) =_{p+1}\!F_{p}\left(
a_{0},\;a^{p},\;b^{p},\;z\right), 
\]
and 
\begin{eqnarray}
&&u_{j}^{p+1}\left( a_{0},\;a^{p},\;b^{p},\;z\right)  \nonumber \\
&=&z^{1-b_{j}}\,_{p+1}\!F_{p}\left(
a_{0}-b_{j}+1,\;a_{i}-b_{j}+1,\;1+b_{i}-b_{j},\;2-b_{j};\;z\right)  
\end{eqnarray}

Since the solution we are looking for, has to be monovalued in $z\bar{z}$ 
the $\lambda _{ij}\,^{\prime }{\rm s}$ have to be diagonal provided none of
the difference $b_{i}-b_{j}$ is an integer. Thus the general solution reduces to

\begin{eqnarray}
&&I_{p+1}\left( a_{0},\;a^{p},\;b^{p};\;\tilde{a}_{0},\;\tilde{a}^{p},\;%
\tilde{b}^{p};\;z,\bar{z}\right)  \nonumber \\
&=&\sum\limits_{j=0}^{p}\lambda _{j}^{p+1}\left( a_{0},\;a^{p},\;b^{p};\;%
\tilde{a}_{0},\;\tilde{a}^{p},\;\tilde{b}^{p}\right) u_{j}^{p+1}\left(
a_{0},\;a^{p},\;b^{p};\;z\right) \widetilde{u}_{j}^{p+1}\left( \tilde{a}%
_{0},\;\tilde{a}^{p},\;\tilde{b}^{p};\;\bar{z}\right)  \nonumber \\
&&\;  
\end{eqnarray}

\noindent which exhibits a conformal block structure.

The $p+1$ unknown constants $\lambda _{j}\left( a_{0},\;a^{p},\;b^{p},\;%
\tilde{a}_{0},\;\tilde{a}^{p},\;\tilde{b}^{p}\right) $ are readily obtained
by identifying the different behaviour of the solution $I_{p+1}$ near $z=0.$

At this stage we consider a generalized Euler function ([4]),

\[
B_{\alpha ,\;\beta ;\;\tilde{\alpha},\;\tilde{\beta}}=\frac{1}{2i}
\int d^2t\;t^{\alpha -1}\left( 1-t\right) ^{\beta -1}\left( \bar{t}\right) ^{%
\tilde{\alpha}-1}\left( 1-\bar{t}\right) ^{\tilde{\beta}-1} 
\]

\noindent which is defined and analytic for Re$(\alpha +\tilde\alpha)>0$,
Re$(\beta+\tilde\beta)>0$, and Re$(\alpha +\tilde\alpha +\beta+\tilde\beta)<2$. 
Using arguments similar to those given in the previous section the above 
integral can be evaluated  so that  for $\alpha -\tilde{\alpha},$ $\beta -\tilde{\beta}$ integer,

\begin{equation}
B_{\alpha ,\;\beta ;\;\tilde{\alpha},\;\tilde{\beta}}=\frac{\Gamma \left(
\alpha \right) \Gamma \left( \beta \right) }{\Gamma \left( \alpha +\beta
\right) }\;\frac{\Gamma \left( \tilde{\alpha}\right) \Gamma \left( \tilde{%
\beta}\right) }{\Gamma \left( \tilde{\alpha}+\tilde{\beta}\right) }\;\frac{%
\sin \pi \tilde{\alpha}\;\sin \pi \tilde{\beta}}{\sin \pi \left( \tilde{%
\alpha}+\tilde{\beta}\right) }.  
\end{equation}
\noindent Note that since $\alpha -\tilde{\alpha}$ and $\beta -\tilde{\beta}$ are
integer,

\[
\frac{\sin \pi \tilde{\alpha}\;\sin \pi \tilde{\beta}}{\sin \pi \left( 
\tilde{\alpha}+\tilde{\beta}\right) }\equiv \frac{\sin \pi \alpha \;\sin \pi
\beta }{\sin \pi \left( \alpha +\beta \right) } 
\]

To find $\lambda _{0}^{p+1},$ set $z=0$ in (4) in which case the integral is
readily obtained since it factorizes.

Thus,

\begin{equation}
\lambda _{0}^{p+1}\left( a_{0},\;a^{p},\;b^{p},\;\tilde{a}_{0},\;\tilde{a}%
^{p},\;\tilde{b}^{p}\right) =\tprod\limits_{i=1}^{p}B_{a_{i},\;b_{i}-a_{i};\;%
\tilde{a}_{i},\;\tilde{b}_{i}-\tilde{a}_{i}}  
\end{equation}

\noindent which is $a_{0},\;\tilde{a}_{0}$ independent.

To find $\lambda _{j}^{p+1},$ set
$z_{j}=\left[ \tprod\limits_{k=1,k\ne j}^{p}z_{k}z\right]
^{-1}t$ and observe that

\begin{eqnarray*}
&&dz_{j}\;\left( z_{j}\right) ^{a_{j}-1}\left( 1-z_{j}\right)
^{b_{j}-a_{j}-1}\left( 1-\left( \tprod\limits_{i=1}^{p}z_{i}\right) z\right)
^{-a_{0}} \\
&=&z^{1-b_{j}}\left( dt\right) \;t^{a_{j}-1}\left( 1-t\right)
^{-a_{0}}\left( \left( \tprod\limits_{k=1, k\ne j}^{p}z_{k}\right) z-t\right)
^{b_{j}-a_{j}-1}\tprod\limits_{k=1, k\ne j}^{p}\left( z_{k}\right) ^{1-b_{j}}.
\end{eqnarray*}

In a neighborhood of $z=0$ we find
\begin{eqnarray*}
I_{p+1} &\sim &z^{1-b_{j}}\left( \bar{z}\right) ^{1-\tilde{b}_{j}}\left( 
\frac{1}{2i}\right) ^{p}\int d^2t\;t^{b_{j}-2}\left( 1-t\right) ^{-a_{0}}\;\bar{t}^{\widetilde{b}_{j}-2}\left( 1-\overline{t}\right) ^{-\widetilde{a%
}_{0}} \\
&&\tprod\limits_{k=1, k\ne j}^{p}\;d^2z_{k}\left( z_{k}\right)
^{a_{k}-b_{j}}\left( 1-z_{k}\right) ^{b_{k}-a_{k}-1}\left( \bar{z%
}_{k}\right) ^{\widetilde{a}_{k}-\widetilde{b}_{j}} \\
&&\left( 1-\bar{z}_{k}\right) ^{\widetilde{b}_{k}-\widetilde{a}_{k}-1}\left(
-1\right) ^{b_{j}-\tilde{b}_{j}-a_{j}+\tilde{a}_{j}}
\end{eqnarray*}

Thus,

\begin{eqnarray}
\lambda _{j}^{p+1}\left( a_{0},a^{p},b^{p},\widetilde{a}_{0},\widetilde{a}%
^{p},\widetilde{b}^{p}\right) &=&\left( -1\right) ^{b_{j}-\widetilde{b}%
_{j}-a_{j}+\widetilde{a}_{j}}B_{b_{j}-1,\;-a_{0}+1;\widetilde{\;b}_{j}-1,\;-%
\widetilde{a}_{0}+1}\times  \nonumber \\
&&\times \tprod\limits_{i=1, i\ne j}^{p}B_{a_{i}-b_{j}+1,\;b_{i}-a_{i};%
\widetilde{\;a}_{i}-\widetilde{b}_{j}+1,\widetilde{\;b}_{i}-\widetilde{a}%
_{i}}  
\end{eqnarray}

We now give some lemmas which will help prove that $I_{p+1}$ has the 
form indicated above. 
 
Define,
$$
H_{j}^{p}\left(
a_{0},\;a^{p},\;b^{p},\;z,\;z_{p}\right) =\left( z_{p}\right)
^{a_{p}-1}\left( 1-z_{p}\right) ^{b_{p}-a_{p}-1}u_{j}^{p}\left(
a_{0},\;a^{p-1},\;b^{p-1},\;zz_{p}\right), $$
and
$$
K_{j}^{p}\left(
a_{0},\;a^{p},\;b^{p},\;z,\;z_{p}\right) =\left( z_{p}\right) ^{a_{p}-1}\left(
1-z_{p}\right) ^{b_{p}-a_{p}+1}u_{j}^{p}\left(
a_{0},\;a^{p-1},\;b^{p-1},\;zz_{p}\right). 
$$

\begin{lemma} Suppose condition {\bf C} is satisfied.  Then,
\begin{eqnarray}
O_{z}^{p+1}\;H_{0}^{\left( p\right) }\left(
a_{0},\;a^{p},\;b^{p},\;z,\;z_{p}\right)
=-\tprod\limits_{i=0}^{p-1}\;\left( a_{i}\right) \frac{\partial }{\partial
z_{p}}K_{0}^{p}\left( a_{0}+1,\;a^{p}+1,\;b^{p},\;z,\;z_{p}\right),\  
\end{eqnarray}
and for $ j = 1\ldots p-1$,

\begin{eqnarray}
&&O_{z}^{p+1}\;H^{(p)}_{j}\left(
a_{0},\;a^{p},\;b^{p},\;z,\;z_{p}\right)  \\ \nonumber
&=&-\tprod\limits_{i=0}^{p-1}\;\left( a_{i}-b_{j}+1\right) \frac{%
\partial }{\partial z_{p}}K^{p}_{j}\left(
a_{0}+1,\;a^{p}+1,\;b^{p},\;z,\;z_{p}\right) 
\end{eqnarray}
\end{lemma}
\noindent Proof. Since 
\begin{eqnarray*}
\left( z\frac{\partial}{\partial z}+a_{i}\right)\,_{p+1}\!
F_{p}\left(a_{i},\;b_{i},\;z\right) = a_i \,_{p+1}\!
F_{p}\left(a_{i}+1,\;b_{i},\;z\right),
\end{eqnarray*}
 $$\left( z\frac{\partial}{\partial z}+b_{i}-1\right)\,_{p+1}\!F_{p}
\left( a_{i},\;b_{i},\;z\right) = (b_i-1)\,_{p+1}\!F_{p}
\left( a_{i},\;b_{i}-1,\;z\right),$$ and 
$$\frac{\partial}{\partial z}\,_{p+1}\!F_{p}\left( a_{i},\;b_{i},\;z\right)= 
a_0\tprod\limits_{k=1}^{p}{a_k\over b_k}\,_{p+1}\!F_{p}
\left(a_{i}+1,\;b_{i}+1,\;z\right),$$ 
equation (12) now follows after some algebra. Equation (13) similarly follows 
from the identities,
\begin{eqnarray*}
&&\left( z\frac{\partial}{\partial z}+a_{i}\right)z^{1-b_j}\,_{p+1}\!F_{p}
\left( a_{i}-b_j+1,\;1+b_{i}-b_j,2-b_j\;z\right)\\ & 
= &(a_i-b_j+1)z^{1-b_j} \,_{p+1}\!F_{p}\left( a_{i}-b_j+2,\;1+b_{i}-b_j,2-b_j,\;z\right)
\end{eqnarray*}
\begin{eqnarray*}
&&\left( z\frac{\partial}{\partial z}+b_{i}-1\right)z^{1-b_j}\,_{p+1}\!F_{p}
\left( a_{i}-b_j+1,\;1+b_{i}-b_j,2-b_j,\;z\right)\\ 
&=& (b_i-b_j)z^{1-b_j}\,_{p+1}\!F_{p}\left( a_{i}-b_j+1,\;b_{i}-b_j,2-b_j\;z\right),
\end{eqnarray*}
\begin{eqnarray*}
&&\left( z\frac{\partial}{\partial z}+b_{j}-1\right)z^{1-b_j}\,_{p+1}\!
F_{p}\left( (a_{i}-b_j+1,\;1+b_{i}-b_j,2-b_j,\;z\right)\\ 
& =&\left ({(a_0-
b_j+1)\prod_{k=1}^p(a_i-b_j+1)\over (2-b_j)\prod_{k=1}^p
(b_i-b_j+1)}\right)z^{2-b_j}\,_{p+1}\!F_{p}\left( a_{i}-b_j+2,\;
b_{i}-b_j+2,3-b_j\;z\right),
\end{eqnarray*}
 and 
\begin{eqnarray*}
&&\frac{\partial}{\partial z}z^{1-b_j}\,_{p+1}\!F_{p}\left(a_{i}-b_j+1,\;
b_{i}-b_j+1,2-b_j,\;z\right)\\ &=& (1-b_j)z^{-b_j}\,_{p+1}\!F_{p}
\left( a_{i}-b_j+1,\;b_{i}-b_j+1,1-b_j\;z\right),
\end{eqnarray*}

Let $\hat I_{p+1}$ be given
for $0<|z|<1$ by equations (8), (10), and (11) where we assume that $a^p$ and $b^p$ satisfy condition $C$. Since for $0<|z|<1$ the hypergeometric functions in$\hat I_{p+1}$ are defined by their series representation  we seek a continuation of $\hat I_{p+1}$
to $|z|>1$.  It follows from the Mellin-Barnes representation for
${}_{p+1}\! F_p (a_0, a^p, b^p,z)$ (Luke [10] p.~149) that 
for $0<{\rm arg\; }z<2\pi$ 
\begin{eqnarray}
{}_{p+1}\! F_p(a_0,a^p,b^p,z)
& =&\frac{\prod\limits^p_{i=1}\Gamma (b_i)}
{\prod\limits^p_{i=0} \Gamma(a_i)}
\sum^p_{j=0}\Gamma (a_j)\frac{\prod\limits^p_{k=0\atop k\ne j}
\Gamma (a_k-a_j)}{\prod\limits^p_{k=1}\Gamma (b_k-a_j)}
\left(\frac{e^{i\pi}}z\right)^{a_j}\nonumber\\
 &&\times\; {}_{p+1}\!F_p \left(a_j,
1+a_j-b_i, 1+a_j-a_i, \frac1z\right)\end{eqnarray}

The following Lemma will be useful it what follows:

\begin{lemma} Suppose the points $x_i$ and $y_j$, $i=0\ldots p,\ j=1,\ldots
p$ are distinct and none are equal to one. Then,

\begin{eqnarray}
B_{n,k}&=&\prod^p_{i=1}
\frac{1-x_i}{1-y_i}\nonumber\\
&& + 
\sum^p_{j=1}\frac{(1-x_n)(1-x_k)(x_0-y_j)(y_j-x_j)}{(1-y_j)(1-x_0)(y_j-x_n)(y_j-x_k)}
\prod^p_{i=1\atop i\ne j}
\frac{(x_i-y_j)}{(y_i-y_j)}
\nonumber\\
&=& \frac{u_n\prod^p_{i=0\atop i\ne
  j}(u_j-u_i)}{u_0\prod^p_{i=1}(u_j-z_i)}\delta_{n,j}
\end{eqnarray}
where $z_k=y_k-1 $ and $u_i=x_i-1$.
\end{lemma}

\noindent Proof.
$B_{n,k}$ is equal to
$$B_{n,k}=\prod^p_{i=1}
\frac{u_i}{z_i}+
\sum^p_{j=1}\frac{u_n u_k(u_0-z_j)(z_j-u_j)}
{z_j u_0(z_j-u_n)(z_j-u_k)}
\prod^p_{i=1\atop i\ne j}
\frac{(u_i-z_j)}{(z_i-z_j)}
$$
Write $B_{n,k}\prod^p_1\frac{z_i}{u_i} = S_{n,k}$ where, 
$$
S_{n,k}=1
- \lambda
\sum^p_{j=1}\frac{(z_j-u_0)(z_j-u_j)}
{z_j(z_j-u_n)(z_j-u_k)}
\prod^p_{i=1\atop i\ne j}
\frac{(u_i-z_j)}{(z_i-z_j)},
$$
and $\lambda = {u_k u_n\over u_0}\prod_{i=1}^p{z_i\over u_i}$. The
function 
$$f(z)=\lambda\frac{(z-u_0)}{z(z-u_n)(z-u_k)}\prod_{i=1}^p\frac{(z-u_i)}{(z-z_i)}$$
behaves as ${z}^{-2}$ for large z,
and by hypothesis has simple poles at $z=z_j$ and $z=0$. If $n=k$ then
there is one extra simple pole at $z=u_k$. 
Cauchy's theorem says the sum of the residues is zero and the evaluation of
the residues gives the result.
 
With $s(a_i) = \sin\pi a_i$ we now prove,

\begin{lemma} Suppose $a_0$, $\tilde a_0$, $a^p$, 
 $\tilde a^p$, $\tilde b^p$ and $b^p$ satisfy condition {\bf C} then
\end{lemma}

\begin{eqnarray*}
\hat I_{p+1}(z,\bar z)&=&\sum^p_{j=0} J_{jj}e^{i\pi (a_j-\tilde a_j)}
z^{-a_j} \bar z^{-\tilde a_j} 
{}_{p+1}\!F_p\left(a_j,a_j-b_i+1,1+a_j-a_{l\ne j},\frac1z\right)\\
&& \times\; {}_{p+1}\!F_p \left(\tilde a_j, \tilde a_j-\tilde b_i+1,
1+\tilde a_j-\tilde a_{l\ne j},\frac1{\bar z}\right)
\end{eqnarray*}
where $J_{jj}$ is given by,
\begin{eqnarray}
J_{jj}=\frac{s(\tilde b_j-\tilde a_j)}{s(\tilde a_0-\tilde a_j)}B_{a_j,a_0-a_j,\tilde a_j,\tilde a_0-a_j}\prod^p_{i=0 \atop i\ne j}B_{b_i-a_i,a_i-a_j,\tilde b_i-\tilde a_i,\tilde a_i-\tilde a_j}, j\ne 0
\end{eqnarray}
and
\begin{eqnarray}
J_{00}=\prod^p_{i=1}B_{b_i-a_i,a_i-a_0,\tilde b_i-\tilde a_i,\tilde a_i-\tilde a_0}.
\end{eqnarray}
\noindent Proof. From equation (14) it follows that
\begin{eqnarray*}
&&\lambda^{p+1}_0\; {}_{p+1}\!F_p(a_0, a^p,b^p,z)\, {}_{p+1}F_p(\tilde a_0, 
\tilde a^p, \tilde b^p;\bar z)
 =\prod^p_{i=1} \frac{s(\tilde a_i)
s(\tilde b_i-\tilde a _i)}{s(\tilde b_i)}\\
&&\quad \times \left(\sum^p_{j=0} c^{p}_{0j}\left( \frac{e^{i\pi}}z\right)^{a_j}
{}_{p+1}\!F_p\left(a_j,1+a_j-b_i, 1+a_j-a_{l\ne j}; \frac1z\right)\right)\\
&&\quad\times\; \Biggl(\sum^p_{j=0} 
\tilde c^{p}_{0j}\left(\frac{e^{-i\pi}}{\bar z}
\right)^{\tilde a_j} {}_{p+1}\!F_p\left(\tilde a_j, 1+\tilde a_j-\tilde b_i,
1+\tilde a_j-\tilde a_{l\ne j};\frac1{\bar z}\right)\Biggr),
\end{eqnarray*}
where
$$c^{p}_{00} = \prod^p_{i=1} B(b_i-a_i,a_i-a_0),$$
and
$$c^{p}_{0j}=B(a_j,a_0-a_j)\prod^p_{i=1\atop i\ne j}
B(b_i-a_i,a_i-a_j),$$
with $B(a,b)=\frac{\Gamma (a)\Gamma (b)}{\Gamma (a+b)}$.
$\tilde c^{p}_{0k}$ is obtained from $c^{p}_{0k}$ by replacing $a_0$, $a^p$
and $b^p$ by $\tilde a_0$, $\tilde a^p$ and $\tilde b^p$ respectively.
Likewise we find
\begin{eqnarray*}
&&\sum^p_{j=1} \lambda^{p+1}_j z^{1-b_J}\bar z^{1-\tilde b_j}{}_{p+1}\!F_p(a_l-b_j+1,
1+b_i-b_j, 2-b_j,z)\\
&&\times\; {}_{p+1}\!F_p\left(\tilde a_l-\tilde b_j+1,
1+\tilde b_i-\tilde b_j, 2-\tilde b_j; \frac1{\bar z}\right)\\
&&\quad = \sum^p_{j=1} (-1)^{a_j-\tilde a_j}
\frac{s(\tilde b_j-1)s(1-\tilde a_0)}{s(\tilde b_j-
\tilde a_0)}\\
&&\qquad\times
\prod^p_{i=1\atop i\ne j}\frac{s(\tilde a_i-\tilde b_j+1)
s(\tilde b_i-\tilde a_j)}{s(\tilde b_i-\tilde b_j+1)}\\
&&\qquad\times \left(\sum^p_{k=0}c^{p}_{jk}\left(\frac{e^{i\pi}}z\right)^{a_k}
{}_{p+1}\!F_p\left(a_k,1+a_k-b_i,1+a_k-a_{l\ne k};\frac1z\right)\right)\\
&&\qquad\times\left(\left(\sum^p_{k=0}\tilde c^{p}_{jk}
\left(\frac{e^{-i\pi}}{\bar z}\right)^{\tilde a_k}
{}_{p+1}\!F_p (\tilde a_k,1+\tilde a_k-\tilde b_i
1+\tilde a_k-\tilde a_{l\ne k};\frac1{\bar z}\right)\right),
\end{eqnarray*}
with 
$$c^{p}_{j0} = \frac{\Gamma (2-b_j)\Gamma (b_j-1)}{\Gamma (b_j-a_j)
\Gamma(a_j-b_j+1)}\prod^p_{i=1} B(b_i-a_i, a_i-a_0),$$
and
$$c^{p}_{jk}=\frac{\Gamma(2-b_j)\Gamma (b_j-1)}
{\Gamma (b_j-a_0)\Gamma (a_0-b_j+1)} B(1-a_0, a_0-a_k)
\prod^p_{i=1\atop i\ne k} B(b_i-a_i, a_i-a_k)
\frac{s(b_j-a_j)}{s(b_j-a_k)}.$$
With the simplifications
$$\frac{s(\tilde a_i-\tilde b_j+1)}{s(\tilde b_i-\tilde b_j
+1)} = \frac{s(\tilde a_i-\tilde b_j)}{s(\tilde b_i-
\tilde b_j)} ,$$
$$\frac{s( \tilde b_j-1)\Gamma (2-\tilde b_j)\Gamma (\tilde b_j-1)}
{s(\tilde b_j-\tilde a_0)\Gamma (\tilde b_j-\tilde a_0)\Gamma 
(\tilde a_0-\tilde b_j+1)}=1,$$
$$\frac{s(1-\tilde a_0)}{s(b_j-1)}=-\frac{s(\tilde a_0)}
{s(b_j)},$$
and 
$$B(1-a_0,a_0-a_l)=B(a_l,a_0-a_l)\frac{s(a_l)}{s(a_0)},$$
the coefficient $J_{nk},\ n\ne0\ne k$ of
\begin{eqnarray*}
&&\left(\frac{e^{i\pi}}z\right)^{a_n}\left(\frac{e^{-i\pi}}
{\bar z}\right)^{\tilde a_k}{}_{p+1}\!F_p\left(a_n,a_n-b_i+1, 1+a_n-a_{l\ne n}\;frac1z
\right)\\&&\times {}_{p+1}\!F_p\left(\tilde a_k, \tilde a_k-\tilde b_i+1,
1+\tilde a_k-\tilde a_{l\ne k};\frac1{\bar z}\right)
\end{eqnarray*}
becomes,
\begin{eqnarray}
J_{nk}&=&\prod^p_{i=0\atop i\ne n} B(b_i-a_i,a_i-a_n)
\prod^p_{i=0\atop i\ne k}
B(\tilde b_i-\tilde a_i, \tilde a_i-\tilde a_k)\nonumber\\
&&\times \prod^p_{i=1}\sin\pi (\tilde b_i-\tilde a_i)
B(a_n,a_0-a_n)B(\tilde a_k, \tilde a_0-\tilde a_k) \hat J_{nk},
\end{eqnarray}
where
\begin{eqnarray}
\tilde J_{nk} &=& \prod^p_{i=1}\frac{s(\tilde a_i)}{s(\tilde b_i)}
-\sum^p_{j=1}(-1)^{a_j-\tilde a_j}
\frac{s(b_j-a_0)s(a_n)s(\tilde a_k)s(b_j-a_j)}
{s(b_j)s(a _0)s(b_j-a_n)s(\tilde b_j-\tilde a_k)}
\nonumber\\
&&\times\prod^p_{p=1\atop i\ne j}\frac{s(\tilde a_i-\tilde b_j)}
{s(\tilde b_i-\tilde b_j)}.
\end{eqnarray}
Now from the fact that $a_i-\tilde a_i$ and $b_i-\tilde b_i$ differ by
integers we find,
\begin{eqnarray*}
&&(-1)^{a_j-\tilde a_j}\frac{s(b_j-a_0)s(a_n)s(b_j-a_j)}
{s(b_j)s(a_0)s(b_j-a_n)}\\
&&\qquad = -\frac{s(\tilde b_j-\tilde a_0)s(\tilde a_0)
s(\tilde a_n)s(\tilde a_j-\tilde b_j)}
{s(\tilde b_j)s(\tilde a_0)s(\tilde b_j-\tilde a_n)}
\end{eqnarray*}
so that
\begin{eqnarray}
\tilde J_{nk}&=&\prod^p_{i=1}
\frac{s(\tilde a_i)}{s(\tilde b_j)}\nonumber\\
&& + 
\sum^p_{j=1}\frac{s(\tilde b_j-\tilde a_0)s(\tilde a_n)
s(\tilde a_k)s(\tilde a_j-\tilde b_j)}
{s(\tilde b_j)s(\tilde a_0)s(\tilde a_n-\tilde b_j)
s(\tilde a_k-\tilde b_j)}
\prod^p_{i=1\atop i\ne j}
\frac{s (\tilde a_i-\tilde b_j)}{s (\tilde b_i-\tilde b_j)}.
\end{eqnarray}
For $n=0\ne k$ the coefficient $J_{0k}$ of
\begin{eqnarray*}
&&\left(\frac{e^{i\pi}}z\right)^{a_0}\left(\frac{e^{-i\pi}}{\bar z}
\right)^{\tilde a_k}{}_{p+1}\!F_p\left(a_0,1+a_0-b_i,1+a_0-a_i,\frac1z\right)
\\
&&\times\; {}_{p+1}\!F_p
\left(\tilde a_k, 1+\tilde a_k-\tilde b_i, 1+\tilde a_k-
\tilde a_{l\ne k},\frac1z\right)\end{eqnarray*}
becomes, with the substitutions above,
$$J_{0,k} = \prod^p_{i=1}B(b_i-a_i,a_i-a_0)\prod^p_{i=1\atop i\ne k}
B(\tilde b_i-\tilde a_i,\tilde a_i-\tilde a_k)B(\tilde a_k,
\tilde a_0-\tilde a_k)\hat J_{0,k}$$
where $\tilde J_{0,k}$ is given by (18) above with $n=k$.
For $k=0$ we find
\begin{eqnarray}
J_{0,0}&=&\prod^p_{i=1} B(b_i-a_i,a_i-a_0)\prod^p_{i=1} B(\tilde b_i-\tilde a_i,
\tilde a_i-\tilde a_i)\prod^p_{i=1}s (\tilde b_i-\tilde a_i)\nonumber
\\
&&\times\left[\prod^p_{i=1}\frac{s(\tilde a_i)}{s(\tilde b_i)}
-\sum^p_{j=1}\frac{s(\tilde a_0)s(\tilde b_j-\tilde a_j)}{
s(\tilde b_j)s(\tilde b_j-\tilde a_j)}\prod^p_{i=1\atop i\ne j}
\frac{s(\tilde a_i-\tilde b_j)}{s(\tilde b_i-\tilde b_j)}
\right].\end{eqnarray}
With the substitutions $y_j = \exp-2i\pi b_j,\ u_i=\exp-2i\pi a_i$ and $B_{n,k} = J_{n,k}\prod^p_i\frac{\exp{i\pi a_i}}{\exp{i\pi b_i}}$ the result now follows from Lemma 3 and the symmetry between $J_{n,0}$ and $J_{0,n}$.

Define
$$
\psi_p= \sum_{i=1}^{p}b_i - \sum_{i=0}^{p}a_i,
$$
and $\tilde\psi_p$ as above with $b_i$ and $a_i$ replaced by $\tilde b_i$ 
and $\tilde a_i$ respectively.
\begin{lemma} Suppose Condition $C$ is satisfied, $\psi_{p-1}$ is not an integer, ${\rm Re}(\psi_{p-1}+\tilde\psi_{p-1})>p-1$, ${\rm Re}(a_p +\tilde a_p)<0$, ${\rm Re}(b_p + \tilde b_p - a_p + \tilde a_p)<0$, ${\rm Re}(b_j + \tilde b_j - a_p - \tilde a_p )< 2$, and ${\rm Re}(b_p + \tilde b_p - a_j - \tilde a_j) < 2$.  Then for $ z \ne 0$ or $1$
\begin{eqnarray}
&&\mathop{\int}_{\Gamma}d\bar z_p a_0\prod_{i=0}^{p-1} a_i \ K^{p}_0(a_0+1, a^p+1, b^p, z, z_p)H^{p}_0(\tilde a_0, \tilde a^p, \bar z, \bar z_p)\\\nonumber&& +\prod_{i=0}^{p-1}(a_i-b_j+1)\ K^{p}_j(a_0+1, a^p+1, b^p, z, z_p)H^{p}_j(\tilde a_0, \tilde a^p, \tilde b^p, \bar z, \bar z_p) =0,
\end{eqnarray}
where $\Gamma = [0,\infty\exp{-i\phi})\cup[1,\infty\exp{-i\phi})$, 
$\phi =\arg(z)$.
\end{lemma}
\noindent Proof. We use $\Gamma$ a cuts 
for $z^{\alpha}$ and $(1-z)^{\alpha}$ choosing the determination for both 
so that both give positive reals when their arguments a large positive real 
numbers. We beginning by breaking up the above integral into  pieces 
for which $|zz_p|< 1$  and $|zz_p|>1$. For $|zz_p|>1$ we use the fact that
\begin{eqnarray}
&&\prod_{i=0}^{p-1}a_i \lambda^{p-1}_0(a_0, a^{p-1}, b^{p-1}, \tilde a_0, \tilde a^{p-1}, \tilde b^{p-1})\\ &&\nonumber = \prod_{i=1}^{p-1}(b_i-a_i-1)\lambda^{p-1}_0(a_0+1, a^{p-1}+1, b^{p-1}, \tilde a_0, \tilde a^{p-1}, \tilde b^{p-1}),
\end{eqnarray}
and
\begin{eqnarray}
&&\prod_{i=0}^{p-1}(a_i-b_j+1) \lambda^{p-1}_j(a_0, a^{p-1}, b^{p-1}, \tilde a_0, \tilde a^{p-1}, \tilde b^{p-1})\\ &&\nonumber =a_0\prod_{i=1}^{p-1}(b_i-a_i-1) \lambda^{p-1}_j(a_0+1, a^{p-1}+1, b^{p-1}, \tilde a_0+1, \tilde a^{p-1}+1, \tilde b^{p-1}),
\end{eqnarray}
and Lemma~5 to represent the integrand in (22) for $|zz_p|>1$.
The conditions on $a_0 \ldots , a_p$, $b_1 \ldots , a_p$,
 $\tilde a_0 \ldots , \tilde a_p$, and $\tilde b_1 \ldots ,\tilde b_p$ 
imply that the above integral is convergent for $z_p \sim \infty , 0$, and 1.
Furthermore with condition $C$ and the condition on $\psi_{p-1}$ the hypergeometric functions in the integrand behave as 
${}_{p}F_{p-1}(a_0, a^p, b^p, z)\sim (1-z)^{\psi_{p-1}}$ ([8]) so that the integral is convergent for $zz_p \sim 1$. The last two parts of condition $C$ 
insure that the
contributions along the two sides of each cut cancel out giving the result.  

We now prove our main result,
\begin{theorem} Suppose that condition $C$ is satisfied,  $\psi_{p-1}$ is not an integer, ${\rm Re}(\psi_{p-1}+\tilde\psi_{p-1})> -2$, ${\rm Re}(a_p +\tilde a_p)>0$, $ {\rm Re}(b_p + \tilde b_p - a_p - \tilde a_p)>0$, 
${\rm Re}(b_j + \tilde b_j - a_p - \tilde a_p )< 2$, and ${\rm Re}(b_p + \tilde b_p - a_j - \tilde a_j) < 2$. then $I_{p+1}$ is
given by equation (8) with $\lambda^{p+1}_j$ given by (9) and (10). $I_{p+1}$ may be extended using the above representation so that only condition $C$ is 
satisfied by the parameters. 
\end{theorem}
\noindent Proof.  Consider $I_{p+1}$ for $p=1$, the result is already 
known ([4]) and we will recover it. In this case,
\begin{eqnarray}
&&I_2=\\\nonumber \\ &&\mathop{\int\int}_{{\mathbb C}\backslash \Gamma}{d^2z_{1}\over2i}
z_1^{a_1-1}(1-z_1)^{b_1-a_1-1}(1-zz_1)^{-a_0}
\bar{z_1}^{\tilde a_1 -1}(1-\bar z_1)^{\tilde b_1 -\tilde a_1 -1}
(1-\bar z\bar z_1)^{-\tilde a_0}\nonumber \\
&=&\mathop{\int\int}_{{\mathbb C}\backslash \Gamma}{d^2z_{1}\over2i}
z_1^{a_1-1}(1-z_1)^{b_1-a_1-1}\bar{z_1}^{\tilde a_1 -1}(1-\bar z_1)^{\tilde b_1 -\tilde a_1 -1}I_1(a_0,\tilde a_0, zz_1, \bar z \bar z_1)\nonumber\\
&=&\nonumber\mathop{\int\int}_{\mathbb C}{d^2z_{1}\over 2i}\chi_{{\mathbb C}\backslash \Gamma}
z_1^{a_1-1}(1-z_1)^{b_1-a_1-1}\bar{z_1}^{\tilde a_1 -1}(1-\bar z_1)^{\tilde b_1 -\tilde a_1 -1}\\&&\times I_1(a_0,\tilde a_0, zz_1, \bar z \bar z_1).
\end{eqnarray}
where $\Gamma = [0,\infty\exp{-i\phi})\cup[1,\infty\exp{-i\phi})$, 
$\phi =\arg(z)$ and $\chi_{{\mathbb C}\backslash \Gamma}$ is the characteristic function of the set ${{\mathbb C}\backslash \Gamma}$. We will use the determination given in the above lemma 
for $z^{\alpha}$. We suppose that $z \ne 0$ or 1. If Re$(a_1+\tilde a_1)>0$, Re$(b_1+\tilde b_1 - 
a_1 -\tilde a_1)>0$, Re$(a_0+\tilde a_0) < 2$, and 
Re$(b_1+\tilde b_1- a_0-\tilde a_0) < 2$ the above integral converges 
uniformly on compact subsets of the $z$ plane that exclude zero and one and defines an analytic 
function of the $a$ and $b$ variables. Utilizing the third line in the above equation it is not difficult to see that the integral is continuous in $z$. 
For $\epsilon$ small but positive 
let $I_2^{\epsilon}$ be given by the integral above with $\Gamma$ replaced 
by $\Gamma_{\epsilon}$ which is the contour with every point in 
$\Gamma$ displaced by a  distance $\epsilon$. Thus
$$
I_2^{\epsilon} =\mathop{\int\int}_{\mathbb C}d^2z_{1}\chi_{{\mathbb C}\backslash \Gamma_{\epsilon}}
z_1^{a_1-1}(1-z_1)^{b_1-a_1-1}\bar{z_1}^{\tilde a_1 -1}(1-\bar z_1)^{\tilde b_1 -\tilde a_1 -1}I_1(a_0,\tilde a_0, zz_1, \bar z \bar z_1).
$$ 
 The integral converges uniformly for $z$ on compact subsets contained in 
or on  $ \Gamma_{\epsilon}$ that avoid $z=0,1$ and defines an analytic function in the variables $a_0, a_1, b_1, \tilde a_0, \tilde a_1, \tilde b_1$ in the region designated above.  Furthermore  it follows 
 that $I_2^{\epsilon}$
is a continuous function of $\epsilon$ for $z$ sufficiently close to $\Gamma\backslash\{0,1\}$ and $\lim_{\epsilon \to0}I_2^{\epsilon} = I_2$ where the convergence is uniform  on compact subsets of $\Gamma\backslash\{0,1\}$. Since for every $\epsilon$ small but positive 
the integral, and the integral with the integrand differentiated once,  and  the integral with the integrand differentiate twice with respect 
to $z$ converge uniformly, it is possible to interchange integration
and differentiation with respect to $z$. 
  From (12) above we find
\begin{eqnarray*}
O_{z}^{2}\;I_{2}^{\epsilon}\left( z,\;\bar{z}\right)  &=&-a_{0}
\frac{2}{2i}\int d^2z_{1}%
\frac{\partial }{\partial z_{1}}K_{0}^{(1)}\left(
a_{0}+1,\;a_{1}+1,\;b_{1},\;z,\;z_{1}\right)  \\
&&\;H_{0}^{(1)}\left( \widetilde{a}_{0},\;\widetilde{a}_{1},\;%
\tilde{b}_{1},\;\bar{z},\;\bar{z}_{1}\right)
\end{eqnarray*}
Thus from Stokes' Theorem we find 
\begin{eqnarray*}
O_z^2I_2^{\epsilon}&=&{1\over4}\int_{\Gamma_\epsilon} K^{(1)}_0 (z,z_1)H(\bar z,\bar z_1)d\bar z_1\\
&=& G_\epsilon (z,\bar z)\end{eqnarray*}
To continue on we impose the extra restriction that 
${\rm Re}(a_0 + \tilde a_0)<0$. Under these conditions it follows using 
Stokes' Theorem that for $z$ sufficiently close 
to $\Gamma\backslash\{0,1\}, G_\epsilon (z,\bar z)$ is continuous 
in $\epsilon$. From Lemma~6 we also see that 
$\lim_{\epsilon\to0} G_\epsilon (z,\bar z) = 0$ uniformly on compact subsets of $\Gamma\backslash\{0,1\}$.
Thus we can conclude that for  $z$
inside $\Gamma_\epsilon$. 
$$I^2_\epsilon (z)=\psi _h(z)+\psi_p(z)$$
where $\psi_h(z)=c^\epsilon_0 u^2_0 + c^\epsilon_1 u^2_1$ and
$$\psi_p(z)=\sum_{i=0}^1 u^2_i(z)\int^z_{z_0}\frac{w_i(u^2_0, u^2_1)(u)}
{w(u^2_0, u^2_1)(u)}G_\epsilon (u)du$$
where $w(u^2_0,u^2_1)$ is the Wronskian of $u^2_0$ and $u^2_1$ while
$w_i(u^2_0, u^2_1)$ is the determinant obtained from $w(u^2_0, u^2_1)$ by
replacing the $i$th column of $w$ by ${0\choose 1}$. Condition $C$ insures that $u^2_0$ and $u^2_1$ are linearly independent. Here $z_0$ is 
chosen so that $0<|z_0|<1$ and $z_0$ lies on $\Gamma$.  The coefficients
$c^\epsilon_0$ and $c^\epsilon_1$ are functions of $\bar z$ 
(thought of as independent of $z$ for their computation). A similar discussion
interchanging $z, a_0, a_1$, and $b_1$ with  
$\bar z, \tilde a_0,\tilde a_1$, and $\tilde b_1$ respectively shows that
$$
I^{\epsilon}_2 = \sum^1_{i=0,j=0} \beta^{\epsilon}_{i,j}u^2_i(a_0,a_1,b_1,z)u^2_j(\tilde a_0, \tilde a_1, \tilde b_1, \bar z) + f^{\epsilon},
$$
where $f^{\epsilon}$ is a function of 
$u^2_{0,1}(a_0, a_1, b_1, z)$, $u^2_{0,1}(\tilde a_0, \tilde a_1, \tilde b_1, \bar z)$ and their first derivatives. The theory of differential equations 
[9 Theorem 4.1] says that the $\beta^{\epsilon}_{i,j}$ and 
$f^{\epsilon}$ are continuous functions of $\epsilon$ 
for $z\in\Gamma\backslash\{0,1\}$ and $\lim_{\epsilon\to0}f^{\epsilon} =0$ 
uniformly on compact subsets of $\Gamma\backslash\{0,1\}$. Writing
\begin{eqnarray}
&&I_2 =\nonumber\\&& \mathop{\int\int}_{{{\mathbb C}\backslash \Gamma}\atop{|z_1|\le1}}{d^2z_{1}\over2i}
z_1^{a_1-1}(1-z_1)^{b_1-a_1-1}(1-zz_1)^{-a_0}
\bar{z_1}^{\tilde a_1 -1}(1-\bar z_1)^{\tilde b_1 -\tilde a_1 -1}
(1-\bar z\bar z_1)^{-\tilde a_0}\nonumber\\
&+&\mathop{\int\int}_{{{\mathbb C}\backslash \Gamma}\atop{|z_1|>1}}{d^2z_{1}\over2i}
z_1^{a_1-1}(1-z_1)^{b_1-a_1-1}(1-zz_1)^{-a_0}
\bar{z_1}^{\tilde a_1 -1}(1-\bar z_1)^{\tilde b_1 -\tilde a_1 -1}
(1-\bar z\bar z_1)^{-\tilde a_0}\nonumber\\&=& I_1^2+I_2^2.
\end{eqnarray}
With the above constraints on the parameters we see that in a neighborhood 
of $z=0$, $I^1_2\sim K$ where 
$K$ is a constant independent of $z$ and $\bar z$. The change of variables $t=zz_1$ in the second integral yields
$$
I^2_2 = z^{1-b_1}{\bar z}^{1-\tilde b_1}(-1)^{b_1-\tilde b_1 - a_1+\tilde a_1} \mathop{\int\int}_{{{\mathbb C}\backslash[0 \infty)}\atop{|{t\over z}|>1}}{d^2t\over2i}
t^{b_1-2}(1-{z\over t})^{b_1-a_1-1}(1-t)^{-a_0}
\bar{t}^{\tilde b_1-2}(1-{\bar z\over \bar t})^{\tilde b_1 -\tilde a_1 -1}
(1-\bar t)^{-\tilde a_0}.
$$
Consequently the dominated convergence theorem shows that 
$$
\lim_{z\to 0} z^{b_1- 1}{\bar z}^{\tilde b_1 -1} I^2_2 = \lambda^2_1. 
$$
Since 
the only constraints on $b_1$ and $\tilde b_1$ are those give above and 
condition $C$ the above considerations imply  
that $\beta^0_{0,1}=0=\beta^0_{1,0}$ and $\beta^{1,1}=\lambda^2_1$.
We also find from the dominated convergence theorem that for $z$ real $\lim_{z\to 0}I_2 = \lambda^2_0$. From the previous discussion the 
solution reads
\begin{eqnarray*}
&&I^{\epsilon}_{2}\left( a_{0},\;a_{1},\;b_{1},\;\widetilde{a}_{0},\;
\tilde{a}_{1},\;%
\tilde{b}_{1},\;z,\;\bar{z}\right) \\
&=&\lambda _{0}^{(2)}\left( a_{1},\;b_{1};\;\tilde{a}_{1},\;\tilde{b}%
_{1}\right) \;_{2}\!F_{1}\left( a_{0},\;a_{1},\;b_{1};\;z\right)
\;_{2}\!F_{1}\left( \tilde{a}_{0},\;\tilde{a}_{1},\;\tilde{b}_{1};\;\bar{z}%
\right) \\
&&+\lambda _{1}^{(2)}\left( a_{0},\;a_{1},\;b_{1},\;\widetilde{a}_{0},\;%
\tilde{a}_{1},\;\tilde{b}_{1}\right) \left( z\right) ^{1-b_{1}}\left( \bar{z}%
\right) ^{1-\tilde{b}_{1}} \\
&&_{2}\!F_{1}\left( a_{0}-b_{1}+1,\;a_{1}-b_{1}+1,\;2-b_{1};\;z\right) \\
&&_{2}\!F_{1}\left( \widetilde{a}_{0}-\tilde{b}_{1}+1,\;\tilde{a}_{1}-\tilde{%
b}_{1}+1,\;2-\tilde{b}_{1};\;\bar{z}\right) \\
&&\; \\
&&\lambda _{0}^{(2)}\left( a_{1},\;b_{1},\;\tilde{a}_{1},\;\tilde{b}%
_{1}\right) \\
&=&\frac{\Gamma \left( a_{1}\right) \Gamma \left( b_{1}-a_{1}\right) }{%
\Gamma \left( b_{1}\right) }\;\frac{\Gamma \left( \tilde{a}_{1}\right)
\Gamma \left( \tilde{b}_{1}-\tilde{a}_{1}\right) }{\Gamma \left( \tilde{b}%
_{1}\right) }\;\frac{\sin \pi \tilde{a}_{1}\sin \pi \left( \tilde{b}_{1}-%
\tilde{a}_{1}\right) }{\sin \pi \tilde{b}_{1}}
\end{eqnarray*}

and

\begin{eqnarray*}
&&\lambda _{1}^{(2)}\left( a_{0},\;a_{1},\;b_{1},\;\tilde{a}_{0},\;\tilde{a}%
_{1},\;\tilde{b}_{1}\right) \\
&=&\left( -1\right) ^{b_{1}-\tilde{b}_{1}+a_{1}-\tilde{a}_{1}}\frac{\Gamma
\left( b_{1}-1\right) \Gamma \left( -a_{0}+1\right) }{\Gamma \left(
-a_{0}+b_{1}\right) }\;\frac{\Gamma \left( \tilde{b}_{1}-1\right) \Gamma
\left( -\tilde{a}_{0}+1\right) }{\Gamma \left( -\tilde{a}_{0}+\tilde{b}%
_{1}\right) } \\
&&\times \;\frac{\sin \pi b_{1}\;\sin \pi \left( -a_{0}\right) }{\sin \pi
\left( b_{1}-a_{0}\right) }
\end{eqnarray*}

\noindent in agreement with previous results. This proves the result for
$p=1$ for  Re$(a_1+\tilde a_1)>0$, Re$(b_1+\tilde b_1 - 
a_1 -\tilde a_1)>0$, Re$(a_0+\tilde a_0) < 0$, and 
Re$(b_1+\tilde b_1- a_0-\tilde a_0) < 2$. We may now extend $I_2$ using the 
above representation so that
only condition $C$ is satisfied by the parameters.

Suppose now that,

\begin{eqnarray*}
&&I_{p}\left( a_{0},\;a^{p-1},\;b^{p-1},\;\tilde{a}_{0},\;\tilde{a}^{p-1},
\;\tilde{%
b}^{p-1},\;z,\;\bar{z}\right) \\
&=&\dsum\limits_{j=0}^{p-1}\lambda _{j}^{(p)}\left( a_{0},\;a^{p-1},
\;b^{p-1},\;%
\tilde{a}_{0},\;\tilde{a}^{p-1},\;\tilde{b}^{p-1}\right) u_{j}^{p}\left(
a_{0},\;a^{p-1},\;b^{p-1},\;z\right) \times \\
&&\times \;\tilde{u}_{j}^{p}\left( \tilde{a}_{0},\;\tilde{a}^{p-1}\;\tilde{b}%
^{p-1},\;\bar{z}\right),
\end{eqnarray*}
then
\begin{eqnarray*}
&&\;I_{p+1}\left( a_{0},\;a^{p},\;b^{p},\;\tilde{a}_{0},
\;\tilde{a%
}^{p},\;\tilde{b}^{p},\;z,\;\bar{z}\right) \\
&=&\;\dsum\limits_{j=0}^{p-1}\;\lambda _{j}^{(p)}\left( \tilde{a}%
_{0},\;\tilde{a}^{p-1},\;\tilde{b}^{p-1},\;\tilde{a}_{0},\;\tilde{a}^{p-1},\;%
\tilde{b}^{p-1}\right) \times \\
&&\times \frac{1}{2i}\mathop{\int\int}_{{\mathbb C}\backslash \Gamma} d^2z_{1}\left( z_{1}\right) ^{a_{p}-1}\left(
1-z_{1}\right) ^{b_{p}-a_{p}-1}u_{j}^{p}\left(
a_{0},\;a_{i},\;b_{i},\;zz_{1}\right) \\
&&\left( \bar{z}_{1}\right) ^{\tilde{a}_{p}-1}\left( 1-\bar{z}%
_{1}\right) ^{\tilde{b}_{p}-\tilde{a}_{p}-1}u_{j}^{p}\left( \tilde{a}_{0},\;%
\tilde{a}^{p-1},\;\tilde{b}^{p-1},\;\bar{z}\bar{z}_{1}\right).
\end{eqnarray*}
The conditions imposed above imply that $I_{p+1}$ is analytic in the 
parameters $a_0,a^p, b^p$, $\tilde a_0, \tilde a^p$, and $\tilde b^p$ in 
the region specified. Furthermore $I_{p+1}$ is a continuous function of
$z$ for $z\ne 0,1$. We now impose the condition that 
${\rm Re}(\psi_{p-1}+ \tilde\psi_{p-1})>p-1$. Let $\Gamma^{\epsilon}$ be 
as above and set
\begin{eqnarray*}
&&I^{\epsilon}_{p+1} = \\&&\mathop{\int\int}_{\mathbb C}d^2z_{1}\chi_{{\mathbb C}\backslash \Gamma_{\epsilon}}
z_1^{a_p-1}(1-z_1)^{b_p-a_p-1}\bar{z_1}^{\tilde a_p -1}(1-\bar z_1)^{\tilde b_p -\tilde a_p -1}\\&&\times I_p(a_0, a^{p-1}, b^{p-1} \tilde a_0, \tilde a^{p-1}, \tilde b^{p-1} zz_1, \bar z \bar z_1),
\end{eqnarray*}
and a discussion similar to that given above shows that 
$\lim_{\epsilon\to0}I^{\epsilon}_{p+1} = I_{p+1}$ 
uniformly on compact subsets of $\Gamma\backslash\{0,1\}$. Since $I^{\epsilon}_{p+1}$ can be differentiated twice we find using Stokes' Theorem  that

\begin{eqnarray*}
&&O_{z}^{p+1}\;I_{p+1}\left( a_{i},\;b_{i},\;\tilde{a}_{i},\;b_{i},\;a_{0},\;%
\tilde{a}_{0},\;z,\;\bar{z}\right) \\
&=&-a_{0}\;\tprod\limits_{i=1}^{p-1}\;\left( b_{i}-a_{i}-1\right) \frac{1}{2i%
}\mathop{\int\int}_{{\mathbb C}\backslash\Gamma_{\epsilon}} dz_{1}\;\frac{\partial }{\partial z_{1}}\;z_{1}^{a_{p}}\left(
1-z_{1}\right) ^{b_{p}-a_{p}} \\
&&I_{p}\left( a_{0}+1,\;a_{i}+1,\;b_{i},\;\tilde{a}_{0},\;\tilde{a}_{i},\;%
\tilde{b}_{i},\;zz_{1},\;z\bar{z}_{1}\right) d\bar{z}_{1}\;\bar{z}%
_{p}^{a_{p}-1}\left( 1-\bar{z}_{1}\right) ^{\tilde{b}_{p}-\tilde{a}_{p}-1}\\ &=&-{a_0\prod_{i=1}^{p-1}(b_i-a_i-1)\over 4}\mathop{\int}_{\Gamma_{\epsilon}}d\bar z_1\ K^{p}_0(a_0+1, a^p+1, b^p, z, z_1)H^{p}_0(\tilde a_0, \tilde a^p, \bar z, \bar z_1)\\\nonumber&& +\ K^{p}_j(a_0+1, a^p+1, b^p, z, z_1)H^{p}_j(\tilde a_0, \tilde a^p, \tilde b^p, \bar z, \bar z_1)\\
&=& G^{\epsilon}_{p+1}.
\end{eqnarray*}
Lemma~5 shows that the last integral is convergent and 
$\lim_{\epsilon\to0}G^{\epsilon}_{p+1}=0$ uniformly on compact subsets of
$\Gamma\backslash\{0,1\}$ by Lemma~6. A similar equation is obtained
when using $O^{p+1}_{\bar z}$ and we are led to the solution
$$
I^{\epsilon}_{p+1} = \sum_{i,j =0}^{p+1}\beta_{i,j}^{\epsilon}u_{i}^{p+1}(a_0, a^p, b^p,z)u_{j}^{p+1}(\tilde a_0, \tilde a^p, \tilde b^p \bar z)+ f^{\epsilon},
$$
where condition $C$ insures that the $u_j^{p+1}, j= 0,\ldots, p $ are 
linearly independent.
We again appeal to the theory of differential equations to show that
$\lim_{\epsilon\to0}f^{\epsilon} =0$ and that $\beta_{i,j}^{\epsilon}$ is a continuous function of $\epsilon$. Using the above continuity properties we find
that $I_{p+1} = \sum_{i,j =0}^{p+1}\beta_{i,j}u_{i}^{p+1}u_{j}^{p+1}$. Write
$I_{p+1} = \sum_{j=0}^p I_{p+1}^j$, where,
\begin{eqnarray*}
I_{p+1}^j = {\lambda^{p}_j \over 2i}\mathop{\int\int}_{{\mathbb C}\backslash\Gamma}&&d^2z_1\chi_{|zz_1|<1}(z_1) z_1^{a_p-1}{\bar z_1}^{\tilde a_p-1}(1-z_1)^{b_p-a_p-1}(1-{\bar z_1})^{\tilde b_p-\tilde a_p-1}\\ && \times u^{p}_j(a_0, a^{p-1}, b^{p-1}, zz_1)u^{p}_j(\tilde a_0, \tilde a^{p-1},\tilde  b^{p-1}, \bar z{\bar z_1}),
\end{eqnarray*}
 for $j = 0,\ldots, p-1$, and
\begin{eqnarray*}
&&I_{p+1}^p =\\&& \sum _{j=0}^{p-1}{J_{jj}e^{i\pi(a_j-\tilde a_j)}\over 2i}\mathop{\int\int}_{{\mathbb C}\backslash\Gamma\atop|zz_1>1}d^2z_1 z_1^{a_p-1}{\bar z_1}^{\tilde a_p-1}(1-z_1)^{b_p-a_p-1}(1-\bar z_1)^{\tilde b_p-\tilde a_p-1}(zz_1)^{-a_j}(\bar z\bar z_1)^{-\tilde a_j}\\&& \times {}_pF_{p-1}(a_j, a_j-b_i+1, 1+a_j-a_{i\ne j}, {1\over zz_1}){}_pF_{p-1}(\tilde a_j, \tilde a_j-\tilde b_i + 1,1+ \tilde a_j - \tilde a_{i\ne j},{1\over \bar z{\bar z_1}}).
\end{eqnarray*}
  From the dominated convergence theorem we see that
$\lim_{z\to0}I^0_{p+1} = \lambda_0^{p+1}$ and\hfill\break
$\lim_{z\to0}z^{b_j-1}\bar z^{\tilde b_j -1}I^j_{p+1} = \lambda_j^{p+1}$ for $z$ real. 
Furthermore with the change of variables $t=zz_1$ it is not difficult to see that   $\lim_{z\to0}z^{b_p-1}\bar z^{\tilde b_p -1}I^p_{p+1} = K^{p+1}$, with $K^{p+1}$ independent of $z$.  Condition $C$ and the above discussion shows that $\beta_{i,j}=0, i\ne j$ and 
$\beta_{i,i}=\lambda^{p+1}_i, i = 0,\ldots, p-1$. In order to compute $\beta_{p,p}$ split $I_{p+1}$ up into the
pieces $I^a_{p+1}$ for $|z_1|\le1$ and $I^b_{p+1}$ for $|z_1|>1$. With the change of variables
$t=zz_1$ the second integral becomes,
\begin{eqnarray*}
I^b_{p+1} =z^{1-b_p}\bar z^{1-\tilde b_p} \mathop{\int\int}_{{\mathbb C}\backslash[0,\infty)}&&d^2t\chi_{|t|>|z|}(t) t^{2-b_p}{\bar t}^{2-\tilde b_p}({z\over t}-1)^{b_p-a_p-1}({\bar z\over \bar t}-1)^{\tilde b_p-\tilde a_p-1}\\ && \times I_{p}(a_0, a^{p-1}, b^{p-1}, \tilde a_0, \tilde a^{p-1},\tilde b^{p-1}, t, \bar t).
\end{eqnarray*}

Using the dominated convergence theorem we see that,
\begin{eqnarray*}
&&(-1)^{b_p-\tilde b_p-a_p +\tilde a_p}\lim_{z\to 0}z^{b_p-1}\bar z^{\tilde b_p-1}I^b_{p+1}\\& =&\mathop{\int\int}_{{\mathbb C}\backslash[0, \infty)}d^2t t^{b_p-2}{\bar t}^{\tilde b_p-2}I_{p}(a_0, a^{p-1}, b^{p-1}, \tilde a_0, \tilde a^{p-1},\tilde b^{p-1}, t, \bar t).
\end{eqnarray*}

We now show that the above integral is equal to $(-1)^{b_p-\tilde b_p-a_p +\tilde a_p}\lambda^{p+1}_{p}$. Note that the result has been shown for $p=1$. Now
suppose the result is true up to $p$. With the substitution of (5) for $I_p$
we find,
\begin{eqnarray*}
&&(-1)^{b_p-\tilde b_p-a_p +\tilde a_p}\lim_{z\to 0}z^{b_p-1}\bar z^{\tilde b_p-1}I^b_{p+1}\\& =&\mathop{\int\int}_{{\mathbb C}\backslash[0, \infty)}d^2t t^{b_p-2}{\bar t}^{\tilde b_p-2}\mathop{\int\int}_{{\mathbb C}\backslash\Gamma}d^2z z^{a_{p-1}-1}{\bar z}^{\tilde a_{p-1}-1}(1-z)^{b_{p-1}-a_{p-1}-1}(1-{\bar z})^{\tilde b_{p-1}-\tilde a_{p-1}-1}\\&& I_{p-1}(a_0, a^{p-2}, b^{p-2}, \tilde a_0, \tilde a^{p-2},\tilde b^{p-2}, tz, \bar tz).
\end{eqnarray*}
Decompose the second integral into the regions $R_1 = |z|\le1$ and $R_2 = |z|> 1$ then make the change of variables $y=zt$. The integral $K_1$ over $R_1$ becomes,
\begin{eqnarray*}
K_1=&&\mathop{\int\int}_{{\mathbb C}\backslash[0, \infty)}d^2t t^{b_p-2}{\bar t}^{\tilde b_p-2}\mathop{\int\int}_{{\mathbb C}\backslash[0, \infty)}d^2\left({y\over t}\right)\chi_{B}({y\over t}) \left({y\over t}\right)^{a_{p-1}-1}\left({\bar y\over\bar t}\right)^{\tilde a_{p-1}-1}\\&&\left(1-{y\over t}\right)^{b_{p-1}-a_{p-1}-1}\left(1-{\bar y\over \bar t}\right)^{\tilde b_{p-1}-\tilde a_{p-1}-1}I_{p-1}(a_0, a^{p-2}, b^{p-2}, \tilde a_0, \tilde a^{p-2},\tilde b^{p-2}, y, \bar y),
\end{eqnarray*}
where $B$ the closed unit ball in complex plane. Interchanging the order of integration then setting $y=ut$ yields,
\begin{eqnarray}
K_1=&&\mathop{\int\int}_{{\mathbb C}\backslash[0, \infty)}d^2u\chi_B(u) u^{a_{p-1}-b_p}{\bar u}^{\tilde a_{p-1}-\tilde b_p}(1-u)^{b_{p-1}-a_{p-1}-1}(1-{\bar u})^{\tilde b_{p-1}-\tilde a_{p-1}-1}\nonumber\\&&\mathop{\int\int}_{{\mathbb C}\backslash\Gamma}d^2t (ut)^{b_p-2}({\bar u\bar t })^{\tilde b_p-2}I_{p-1}(a_0, a^{p-2}, b^{p-2}, \tilde a_0, \tilde a^{p-2},\tilde b^{p-2}, ut, \bar u\bar t),
\end{eqnarray}
where $\Gamma$ is the cut described at the beginning of the 
Theorem with $\phi = \arg t$. Finally with the change of variable $w=ut$ we arrive at,
\begin{eqnarray}
K_1&=&\mathop{\int\int}_{{\mathbb C}\backslash[0, \infty)\atop |u|\le 1}
d^2u u^{a_{p-1}-b_p-1}{\bar u}^{\tilde a_{p-1}\tilde b_p-1}
(1-u)^{b_{p-1}-a_{p-1}-1}(1-{\bar u})^{\tilde b_{p-1}
-\tilde a_{p-1}-1}\nonumber\\
&&\mathop{\int\int}_{{\mathbb C}\backslash\Gamma}d^2w 
w^{b_p-2}{\bar w }^{\tilde b_p-2}I_{p-1}(a_0, a^{p-2}, b^{p-2},
 \tilde a_0, \tilde a^{p-2},\tilde b^{p-2}, w,\bar w).
\end{eqnarray}
Applying the same operations to the integral over region $R_2$ and 
utilizing the fact that for $|{y\over t}|>1, 
(1-{y\over t})^{b_{p-1}-a_{p-1}-1} = {y\over t}^{b_{p-1}
-a_{p-1}-1}({t\over y}-1)^{b_{p-1}-a_{p-1}-1}$ 
then adding the two integrals gives,
\begin{eqnarray}
&&(-1)^{b_p-\tilde b_p-a_p +\tilde a_p}\lim_{z\to 0}z^{b_p-1}
\bar z^{\tilde b_p-1}I^b_{p+1}\\
\nonumber& =&\mathop{\int\int}_{{\mathbb C}\backslash\Gamma'}
d^2z z^{a_{p-1}-1}{\bar z}^{\tilde a_{p-1}-1}(1-z)^{b_{p-1}
-a_{p-1}-1}(1-{\bar z})^{\tilde b_{p-1}-\tilde a_{p-1}-1}\\
\nonumber&&B_{b_{p}-1,\;-a_{0}+1;\widetilde{\;b}_{p}-1,\;-%
\widetilde{a}_{0}+1}\tprod\limits_{i=1}^{p-2}B_{a_{i}-b_{p}+1,
\;b_{i}-a_{i};%
\widetilde{\;a}_{i}-\widetilde{b}_{p}+1,\widetilde{\;b}_{i} -\widetilde{a}%
_{i}},
\end{eqnarray}
where the induction hypothesis has been. The claim now follows.

We may now extend the result to the case when 
${\rm Re}(\psi_{p-1}+\tilde \psi_{p-1})>-2$. 
The above representation may be used to define $I_{p+1}$ when only 
condition $C$ is imposed on the 
parameters which completes the proof.

A comment is in order. The result of this integral 
for $|z| < 1$ is easier to get 
 by using a new basis namely
  the $_{p+1}G_p(a_i\hbox{;}b_i\hbox{;}z)$,
   which are nothing but the
hypergeometric functions $_{p+1}F_p(a_i\hbox{;}b_i\hbox{;}z)$
up to a multiplicative factor namely 
$$_{p+1}G_p(a_i\hbox{;}b_i\hbox{;}z)=
{\prod_0^p\Gamma(a_i) \over \prod_1^p
\Gamma(b_i)}{_{p+1}F_p(a_i\hbox{;}b_i\hbox{;}z)}$$ 
The basis of the expansion is now $V_j(z)\hbox{,}\tilde V_j(\bar z)$ 
rather than $U_j(z)\tilde U_j(\bar z)$
 where 
 $$V_j(z)=z^{1-b_j}{_{p+1}G_p(a_i-b_j+1\hbox{;}b_k-b_j+1\hbox{;}z)}$$ 
and 
$$\tilde V_j(\bar z)=z^{1-\tilde b_j}{_{p+1}\tilde G_p(\tilde a_i-\tilde b_j+1\hbox{;}\tilde b_k-\tilde b_j+1\hbox{;}\bar z)}$$
In these formulas  $k \ne j=0\hbox{,}\ p $ 
where by convention $b_0=\tilde b_0=1$.

In this basis the solution reads
$$I=\sum_{j=0}^p\mu_jV_j(z)\tilde V_(\bar z)$$ where the $\mu_j$ are
 obtained from the
corresponding values of the $\lambda_j$.

It turns out that an obvious factorization does occur. Namely, by defining
 $$\mu_{p+1} ={1 \over \pi^2}(-1)^{a_0-\tilde a_0}
 \prod_{i=0}^p\Gamma(b_i-a_i)\Gamma(\tilde b_i-\tilde a_i)S(b_i-a_i)$$
we get 
$$ \mu_j
=\mu_{p+1}\ {\prod_{i=0}^pS(b_j-a_i)\over \prod_{i=0 \ne j}^pS(b_j-b_i)}$$
To be more specific
\begin{eqnarray*}
\mu_0&=&\mu_{p+1}\ 
{\prod_{i=0}^pS(a_i)\over \prod_{i=1 }^pS(b_i)}\cr
\mu_j&=&-\mu_{p+1} {\prod_{i=0}^pS(b_j-a_i)\over  
  S(b_j)\prod_{i=1 \ne j}^pS(b_j-b_i)}\end{eqnarray*}

Similarly, for $|z| > 1$ it is useful to expand  the integral as a 
combination of  hypergeometric functions which are defined for 
$|z| > 1$ namely
$$(z)^{-a_j}\  _{p+1}G_p(a_j \hbox{,}a_j-b_i+1 \hbox{;}a_j-a_i+1 
\hbox{;}{1 \over z})$$
or more specifically 
$$W_j(z)=(z)^{-a_j}\  _{p+1}
G_p(a_j \hbox{,}a_j-b_i+1 \hbox{;}a_j-a_i+1 \hbox{;}{1 \over z})$$
the solution reads now 
$$I=\sum_{j=1}^p  \nu_j\ W_j(z)\tilde W_j(\bar z)$$
where the $\nu_j$ have to be determined from the value of the integral 
for $ z \to\infty$.

We proceed  as before  and it turns out that  the same 
factorisation does occur namely 
$$ \nu_j=(-1)^{s-\tilde s}
\mu_{p+1} \ {\prod_{i=0}^pS(b_i-a_i) \over \prod_{i=1 \ne j}^pS(a_i-a_j)}$$ 
 Comparing $\mu_j$ and $\nu_j$ amounts but for a sign to interchange 
   $b_k$ and $a_k$ as expected with the interchange of $z\to {1 \over z}$.
Indeed if in the integrand we change $z_i \to {1 \over z_i}$  
the new integrand $J$ reads 
\begin{eqnarray*}
J&=&(-1)^{s-\tilde s}{z}^{-a_0}{\bar z }^{-\tilde a_0 }
\prod_{i=1}^p{t_i}^{a_0-b_i}(1-t_i)^{b_i-a_i-1}(1-\prod_{i=1}^pt_i 
{z}^{-1})^{-a_0}\\
&& \times \prod_{i=1}^p{\bar t_i}^{\tilde a_0-\tilde b_i}
(1-\bar t_i)^{\tilde b_i-\tilde a_i-1}(1-\prod_{i=1}^p
\bar t_i { \bar z}^{-1})^{-\tilde a_0}\end{eqnarray*}
It is now obvious that the result of the integral for 
$|z| > 1$ is deduced from
 the result for $|z| <1$  by a simple change of the parameters of 
 the hypergeometric functions of argument ${1\over z} $  up to the multiplicative factor$$ (-1)^{s-\tilde s}{z}^{-a_0}{\bar z }^{-\tilde a_0 } $$ 
namely  if we define $\alpha_i$ and $\beta_i$ as the  parameters 
 we get by simple identification
\begin{eqnarray*}
\alpha_0&=&a_0\\
\alpha_i&=&a_0-b_i+1\\
\beta_i&=&a_0-a_i+1\end{eqnarray*}
which corresponds to the parameters of $W_0(z)$
and 
\begin{eqnarray*}
 \alpha_i-\beta_j+1&=&a_j-b_i+1\\
\beta_i-\beta_j+1&=&a_j-a_i+1\\
2-\beta_j&=&a_j-a_0+1\end{eqnarray*}
which corresponds to the parameters of $W_j(z)$. The above argument can be made rigorous using (14).

To summarize 
$$ I_{p+1}=
\mu_{p+1}\sum_{j=0}^p\mu_jV_j(z)\tilde V_j(\bar z) $$
where 
$$V_j(z)=z^{1-b_j}{_{p+1}G_p(a_i-b_j+1;b_k-b_j+1; z)}$$
 $$\mu_{p+1} ={1 \over \pi^2}(-1)^{a_0-\tilde a_0}
 \prod_{i=0}^p\Gamma(b_i-a_i)\Gamma(\tilde b_i-\tilde a_i)S(b_i-a_i)$$
 and 
$$\mu_j=\ {\prod_{i=0}^pS(b_j-a_i)\over \prod_{i=0 \ne j}^pS(b_j-b_i)}$$
 and a similar formula for $|z|>1$.

\subsection*{Acknowledgments}

The authors would like to thank R.~Peschanski, M. Bauer and 
R.~Guida for useful discussions. We would also like to thank
Ph Di Fransesco for discussions leading to the use of Cauchy's Theorem
in Lemma 4. JG would like to thank the Service Physique Theorique at Saclay
for their hospitality during his visit especially P.~Moussa.  
\vfill\eject

\section*{Appendix A}

In this appendix we give an alternative method to calculate (30). Besides
the conditions imposed in the hypothesis of Theorem (3) it will assumed
that $b_p-\tilde b_p -b_i+\tilde b_i \ge 0$, $ i = 0,\ldots, p-1$.
We need to calculate 
$$I=\sum_{i=0}^{p-1}\lambda_i^{p-1}\int t^{b_p -2}U_i(t)
{\bar t}^{\tilde b_p -2} \tilde U_i(\bar t) d^2t$$
where 
$$U_i(t)= t^{1-b_i} {_pF_{p-1}(a_j-b_i+1\hbox{; }b_k-b_i+1 \hbox{; }t)} $$
  where 
$j,k\ne i =0$, $p-1$, and $\lambda_i^{p-1}$ is given in formula 11 and 
by convention $b_0=1$ \par
We split the domain of integration in two parts, and write $I=I_1+I_2$ 
where $I_1$ is the contribution for $|t| <1$ and  $I_2$ for $|t| >1$.
In the first domain the hypergeometric functions are convergent so that
$$I_1=\sum_{i=0}^{p-1}\lambda_i^{p-1}
\mathop{\int\!\!\int}_{\!\!B/\Gamma }
t^{b_p -2}U_i(t){\bar t}^{\tilde b_p -2}\tilde 
U_i(\bar t) d^2t,$$
where $B$ is the unit disk in the complex plane and $\Gamma$ is the contour used in the proof of Theorem 3.
 To calculate $I_2$ , the other form 
of  the solution  which is valid for $|z|>1$ will be used, namely
$$I_2=\sum_{i=0}^{p-1}\nu_i^{p-1}
\mathop{\int\!\!\int}_{\!\!\!B^c/\Gamma} t^{b_p -2}W_i(t)
{\bar t}^{\tilde b_p -2}\tilde 
W_i(\bar t) d^2t$$
where 
$$W_i(t)= t^{-a_i}{_pF_{p-1}(a_i-b_j+1;a_i-a_k+1;{1\over t})} $$ 
with the same convention  for the indices $j$ and $k$\par
By definition 
$$\lambda_i^{p-1}{\prod_{j=0 \ne i}^{p-1}\Gamma(b_j-b_i+1)
\Gamma(\tilde b_j-\tilde b_i+1)\over\prod_{j=0 }^{p-1}\Gamma(a_j-b_i+1)
\Gamma(\tilde a_j-\tilde b_i+1)}=\mu_{p-1}\mu_i^{p-1}$$
$$\nu_i^{p-1}{\prod_{j=0 \ne i}^{p-1}\Gamma(a_i-a_j+1)
\Gamma(\tilde a_i-\tilde a_j+1)\over\prod_{j=0 }^{p-1}\Gamma(a_i-b_j+1)
\Gamma(\tilde a_i-\tilde b_j+1)}=(-1)^{s-\tilde s}\mu_{p-1}\rho_i^{p-1}$$
$$s=\sum_{i=1}^{p-1}b_i-\sum_{i=0}^{p-1}a_i$$
$$\mu_{p-1}=(-1)^{a_0-\tilde a_0}{1\over \pi^2}\prod_{i=0}^{p-1}
\Gamma(b_i-a_i)\Gamma(\tilde b_i-\tilde a_i)S(b_i-a_i)$$
$$\mu_i^{p-1}={\prod_{j=0}^{p-1}S(b_i-a_j)\over\prod_{j=0 \ne i}^{p-1}
S(b_i-b_j)}$$
and 
$$\rho_i^{p-1}={\prod_{j=0}^{p-1}S(b_j-a_i)\over\prod_{j=0 \ne i}^{p-1}
S(a_j-a_i)}$$
For $w$ real and $|w|\le r<1$ set,
$$I_1^i(w)= \mathop{\int\!\!\int}_{\!\!B/\Gamma }(wt)^{b_p -2}
U_i(wt)(\overline{w t})^{\tilde b_p -2}\tilde 
U_i(\overline{w t}) d^2t$$
and expand $U_i(t)$ (resp $\tilde U_i(\bar t)$) as a series $\sum_ n$ 
(resp $\sum_m$). The integration on the azimuthal angle $\phi$ yields at once
$$m=n+b_p-\tilde b_p-b_i+\tilde b_i.$$
The integral over the modulus of $t$ is also straightforward and yields 
an additional factor$${1\over 2}{\Gamma(n+b_p-b_i)\over \Gamma(n+b_p-b_i+1)}(w^2)^{n+b_p-b_i}$$
Collecting all the pieces
we get$$I_1^i(w)=\pi\mu_{p-1}\mu_i^{p-1} w^{2(b_p-b_i-1)}
\hbox{ }{ _{2p+1}G_{2p}^i (w^2)} $$
where  $_{2p+1}G_{2p}^i(w^2)$ is  the unrenormalized  hyper geometric  function the  upper parameters of which are 
$a_j-b_i+1,\tilde a_j-b_i+1+b_p-\tilde b_p,b_p-b_i$
and the lower ones read
$b_k-b_i+1,\tilde b_k-b_i+1+b_p-\tilde b_p,b_p-b_i+1 ,
 1+b_p-\tilde b_p-b_i+\tilde b_i$.
 
Now define
$$ f_R(w)
={\pi}^2\mu_{p-1}
{1 \over 2{i\pi}}\int_{C_R}dt\prod_{j=0}^{p-1}
{\Gamma(1-b_j-t) \Gamma(b_p-\tilde b_p+\tilde a_j+t) \over
\Gamma(1-a_j-t)\Gamma(b_p-\tilde b_p+\tilde b_j+t)}
 {\Gamma(t+b_p-1)\over\Gamma(t+b_p)}\, w^{2(b_p-1+t)},
$$
where $C_R$ is a loop beginning and ending at $+\infty$ and encircling all poles
of $\Gamma(...-t)$ once in the negative direction but none of the poles of
  $\Gamma(...+t)$. 
This yields at once  
$$I_1(w)=f_R(w) $$
 
 The same method  holds  for $ I_2(w)$, $w$ real and $|w|\ge r>1$
 with the integration over the azimuthal angle
 yielding
 $n=m+b_p-\tilde b_p-a_i+\tilde a_i$
 while the integration on the modulus of $t$ gives
$${1\over 2}{\Gamma(\tilde a_i-\tilde b_p+1+m)\over
\Gamma(\tilde a_i-\tilde b_p+2+m)}
(w^2)^{-m-1}+b_{\tilde p}- a_{\tilde i}.$$
 As before $I_2(w)$ is  a complicated sum of $G$ functions.
 
Now define,
 \begin{eqnarray*} f_L(w)
&=&(-1)^{s-\tilde s}\pi^2\mu_{p-1}{1 \over 2{i\pi}}
\int_{C_l}dt\\
&& \prod_{j=0}^{p-1}{\Gamma(1-b_j-t)
\Gamma(b_p-\tilde b_p+\tilde a_j+t) \over
\Gamma(1-a_j-t)\Gamma(b_p-\tilde b_p+\tilde b_j+t)}
 {\Gamma(t+b_p-1)\over\Gamma(t+b_p)}w^{2(b_p-1+t)},
\end{eqnarray*}
where $C_L$ is a loop beginning and ending at $-\infty$ 
and encircling all poles
of $\Gamma(...+t)$ once in the positive direction but none of the poles of
  $\Gamma(...-t)$.
  
  It is easy to check that 
  $$f_L=-I_2(w)+\hbox{ residus at } t= 1-b_p.$$
 Indeed 
  $$\rho_i^{p-1}=(-1)^{s-\tilde s}{\prod_{j=0}^{p-1}S(\tilde b_j-\tilde a_i)
  \over\prod_{j=0 \ne i}^{p-1}S(\tilde a_j-\tilde a_i)}.$$
  But Cauchy theorem implies  that 
  $$f_L(w)=f_R(w),$$ 
  so that in the limit $w\to1$ we find,
$$I =\pi^2\mu_{p-1}\prod_{j=0}^{p-1}{\Gamma(b_p-b_j)
\Gamma(1-\tilde b_p+\tilde a_j) 
\over\Gamma(b_p-a_j)\Gamma(1-\tilde b_p+\tilde b_j)} $$
or equivalently
$$I =\mu_{p-1}\prod_{j=0}^{p-1}{S(b_p-a_j)\over S(b_p-b_j)}
\prod_{j=0}^{p-1}{\Gamma(1+a_j-b_p)\Gamma(1-\tilde b_p+\tilde a_j) 
\over\Gamma(1-b_p+b_j)\Gamma(1+\tilde b_j-\tilde b_p)} $$
At this stage it is worth noticing that
$$\mu_p=\mu_{p-1} \Gamma(b_p-a_p)\Gamma (\tilde b_p-\tilde a_p)S(b_p-a_p),$$
which can be recast as,
$$(-1)^{b_p-\tilde b_p -(a_p-\tilde a_p)}\mu_{p-1}\pi^2=
  {\mu_p\over  
\Gamma(1+a_p-b_p)\Gamma (1+\tilde a_p-\tilde b_p)S(b_p-a_p)}.$$ 
this yields at once 
$$I=(-1)^{b_p-\tilde b_p -(a_p-\tilde a_p)}\lambda_{p+1}^p.$$
This is the expected result.\par
 
As a nice application we may recover the generalized Euler formula,
namely calculate
$$ I=\int t^{a -1}U_0(t){\bar t}^{\tilde a -1}\tilde U_0(\bar t) d^2t$$
where $U_0(t)=(1-t)^{b-1}=_1F_0(b-1,t)$.
If we follow the method presented above
we are led to calculate 
$$ f_R
={1 \over 2{i\pi}}\int_{C_R}dt{\Gamma(-t)
\Gamma(a-\tilde a+1-\tilde b+t) \over
\Gamma(b-t)\Gamma(a-\tilde a+1+t)}
 {\Gamma(t+a)\over\Gamma(t+a+1)}$$since by identification
 $b_1=a-1$, $a_0=1-b$,
 it is easy to get that
 $$f_R={I} S(b)\Gamma(1-b)\Gamma(1-\tilde b){\pi}^2$$
 by picking the pole at $t=-a$, we get at once
 $$f_L={\Gamma (a)\Gamma(1-\tilde a-\tilde b) \over
\Gamma(b+a)\Gamma(1-\tilde a)}$$
collecting all the pieces
the final result reads
$$I=\pi {\Gamma(a)\Gamma(b)
\Gamma(1-\tilde a-\tilde b)\over
\Gamma(1-\tilde b)\Gamma(b+a)\Gamma(1-\tilde a)}$$


\end{document}